\documentclass[journal]{IEEEtran}
\pdfoutput=1
\usepackage{xcolor,soul,framed} 
\usepackage{cite}
\usepackage{amsmath,amssymb,amsfonts}
\usepackage{algorithmic}
\usepackage{graphicx}
\usepackage{textcomp}

\usepackage{comment}
\usepackage{enumitem}
\usepackage{soul}
\usepackage{bbding}
\usepackage{pifont}
\usepackage{wasysym}
\usepackage{subcaption}
\usepackage{rotating}
\usepackage{bm}
\usepackage[normalem]{ulem} 

\begin{document}

\title{Machine Learning on the COVID-19 Pandemic, Human Mobility and Air Quality: A Review}
\author{Md. Mokhlesur Rahman, 
Kamal Chandra Paul, \IEEEmembership{Student Member, IEEE}, 
Md. Amjad Hossain, 
G. G. Md. Nawaz Ali, \IEEEmembership{Member, IEEE},
Md. Shahinoor Rahman, 
and Jean-Claude Thill}
%






\maketitle

\begin{abstract}
The ongoing COVID-19 global pandemic is affecting every facet of human lives (e.g., public health, education, economy, transportation, and the environment). This novel pandemic and citywide implemented lockdown measures are affecting virus transmission, people’s travel patterns, and air quality. Many studies have been conducted to predict the COVID-19 diffusion, assess the impacts of the pandemic on human mobility and air quality, and assess the impacts of lockdown measures on viral spread with a range of Machine Learning (ML) techniques. This review study aims to analyze results from past research to understand the interactions among the COVID-19 pandemic, lockdown measures, human mobility, and air quality. The critical review of prior studies indicates that urban form, people's socioeconomic and physical conditions, social cohesion, and social distancing measures significantly affect human mobility and COVID-19 transmission. During the COVID-19 pandemic, many people are inclined to use private transportation for necessary travel purposes to mitigate coronavirus-related health problems. This review study also noticed that COVID-19 related lockdown measures significantly improve air quality by reducing the concentration of air pollutants, which in turn improves the COVID-19 situation by reducing respiratory-related sickness and deaths of the people. It is argued that ML is a powerful, effective, and robust analytic paradigm to handle complex and wicked problems such as a global pandemic. This study also discusses policy implications, which will be helpful for policy makers to take prompt actions to moderate the severity of the pandemic and improve urban environments by adopting data-driven analytic methods.
\end{abstract}



\section{Introduction}
The Coronavirus disease 2019 (COVID-19) is an ongoing global pandemic and public health crisis that was first reported in Wuhan, China, in December 2019 \cite {wei2020impacts, silva2020covid, carteni2020mobility, ogen2020assessing, he2020short}. Human movement and interactions are significantly affected by COVID-19 pandemic \cite {white2018disease, roy2020characterizing,luca2020deep}. Besides affecting healthful living, this highly infectious disease is influencing every domain of human lives (e.g., mental health, social life, education, economy, global supply chains, production, mobility, energy consumption, environment, and so on) \cite {silva2020covid, samuel2020covid, samuel2020feeling, rahman2020covid, rahman2021socioeconomic, abu2020analysis, loske2020impact, teixeira2020link}. This study aims to summarize the results from selected studies conducted recently using Machine Learning (ML) techniques to portray the interplay between the COVID-19 pandemic, human mobility, and air quality. Moreover, this study will be helpful for policymakers to take immediate actions to mitigate the severity of the pandemic and improve well-being in urban environments.

\begin{figure}[htbp]
    \centering
    \includegraphics[width=3in]{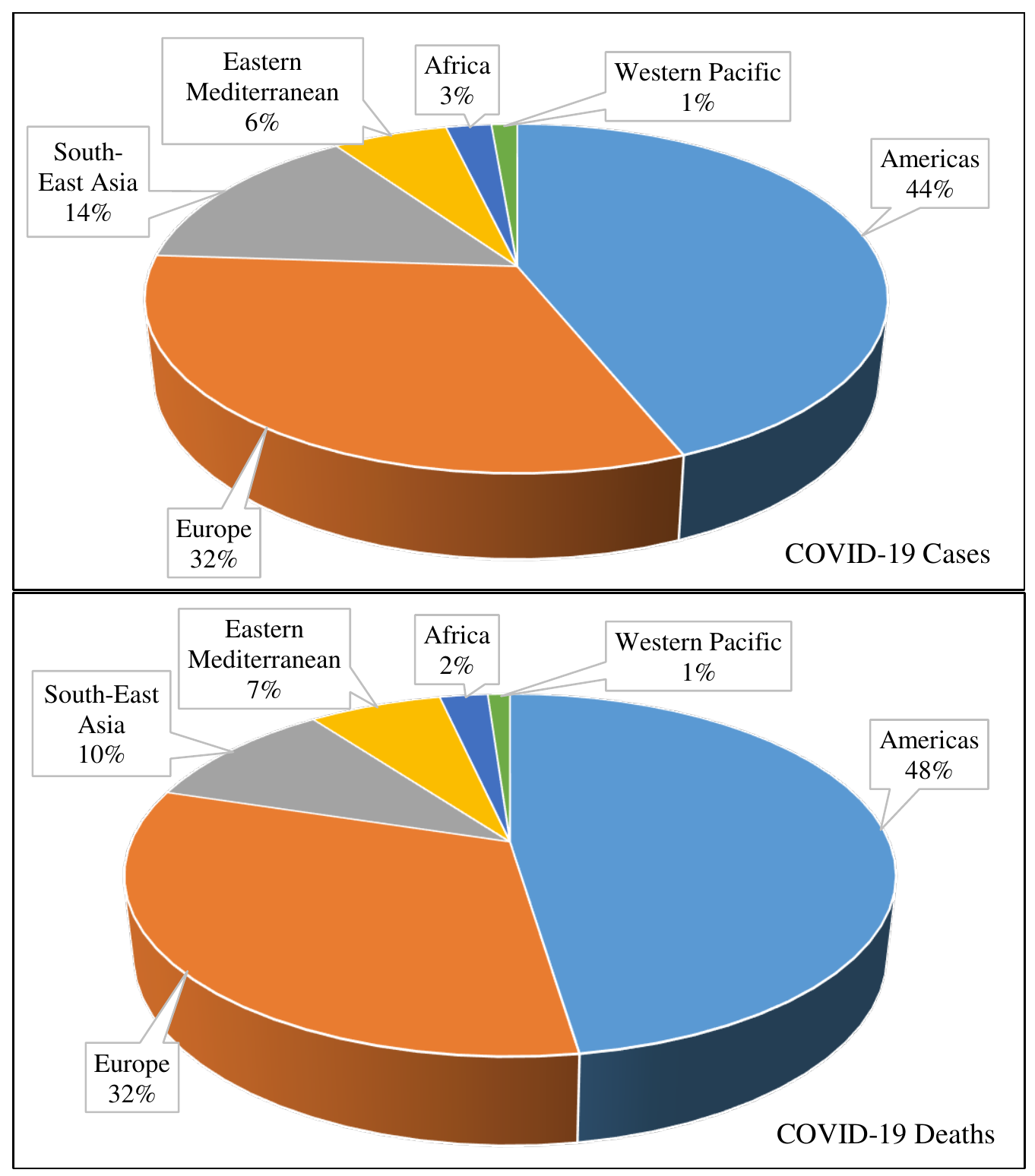}
    \caption{COVID-19 confirmed cases and deaths (\%) by WHO regions (January 6, 2021).}
    \label{Fig:whoregionspie}
\end{figure}

As of January 20, 2021, more than 97 million people from 218 countries have been infected and 2.08 million people have died; this amounted to critical conditions in 0.4\% of active cases and a fatality rate of about 3\% of closed cases \cite {who2020covid19, worldometers2020covid}. Aggregating the total number of confirmed cases and deaths by the World Health Organization (WHO), it is observed that countries in the Americas constitute the most affected region in the world, with about 43.87\% of all confirmed cases and 47.61\% of the deaths (Fig. \ref{Fig:whoregionspie}). Europe is the second most affected region with 32.18\% of confirmed cases and 32.19\% of deaths, followed by South-East Asia with 14.31\% of confirmed cases and 10.03\% of deaths. In contrast, the Western Pacific Region is the least affected with 1.34\% of confirmed cases and 1.11\% of deaths worldwide.

Controlling the outbreak of COVID-19 has become a major concern of governments and public health professionals. Different measures such as stringent travel restrictions and self-quarantine measures (e.g., lockdown, emergency stay-at-home orders, closing of public transportation Systems, travel ban, restrictions on public gathering, closure of workplaces and educational institutions) and personal protective measures (e.g., face masks, gloves) have been implemented \cite{silva2020covid, samuel2020feeling, rahman2020covid, rahman2021socioeconomic, spelta2020after, vinceti2020lockdown, maji2020implication} throughout the world to control the COVID-19 outbreak. These various measures undertaken by governments have influenced both essential (e.g., work) and non-essential (e.g., recreation) trips. For example, lockdown measures undertaken in Singapore to mitigate the COVID-19 pandemic caused a 30\% reduction in mobility \cite {jiang2020spatial}. A 64.6\% reduction in private vehicle trips was observed in Rome, Italy, during lockdown periods of March-April 2020 \cite {aletta2020analysing}. Similarly, about 80\%, 23\%, and 2\% reduction in public transport, cycling and bike-sharing, respectively, were observed in Budapest, Hungary \cite {bucsky2020modal}. In contrast, a 43 to 65\% increase in car travel was reported in Budapest, Hungary, due to the voluntary practice of social distancing and avoidance of public transportation. Air transportation has also been devastated by this COVID-19 situation. Researchers have reported that Canadian Civil Aviation and military aviation activities dropped by 71\% and 27\%, respectively, compared to the business as usual situation \cite {abu2020analysis}. Thus, the widespread diffusion of COVID-19 is adversely affecting all modes of transportation.  

\begin{figure*}[ht]
    \centering
    \includegraphics[width=6.5in]{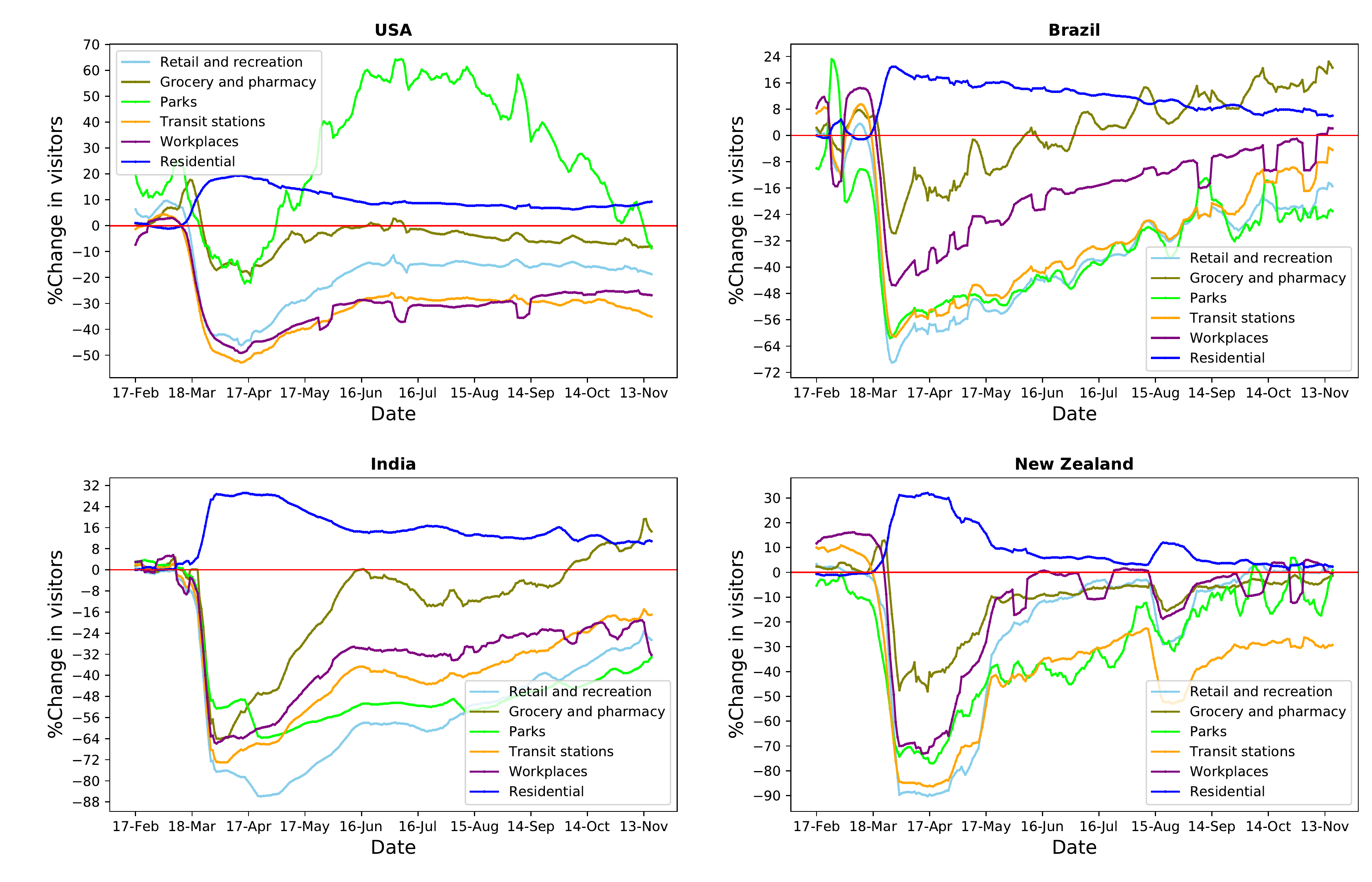}
    \caption{Longitudinal changes in mobility in selected countries.}
    \label{Fig:Changes_mobility}
\end{figure*}

Strict confinement measures have seriously affected mobility in public places and spaces. Fig. \ref{Fig:Changes_mobility} shows these effects in selected countries (US, Brazil, India, and New Zealand) during the period from February 17 to November 27 in 2020. The figure shows a reduction in the number of visitors (\%) in retail and recreation outlets, groceries and pharmacies, parks, transit stations, and workplaces compared to the baseline scenario (i.e., the median value of the day for the 5‑week period starting from January 03 to February 06, 2020), with a deeper reduction in March and April. In contrast, the number of visitors has increased in residential areas due to implemented social distancing measures (i.e., stay-at-home order), which exemplifies the adverse impacts of the pandemic on travel patterns of the people.  

The COVID-19 pandemic also has had substantial impacts on air quality. As a primary method of slowing down the spread of COVID-19, initially, a lot of countries imposed \textit{lockdown} or \textit{confinement} measures to enforce strict social distancing regulations. As a result, businesses and shops were closed, manufacturing activities were either stopped or shrunken while the number of vehicles in cities has declined dramatically \cite {international2020global, dafnomilis2020exploring, han2020assessing}. Therefore, \textit{lockdown} and \textit{confinement} played a critical role in curtailing emissions and in improving overall air quality. Improved air quality refers to the reduction of concentration of criteria pollutants such as $NO_2$, ${SO_2}$, ${PM_{10}}$, ${PM_{2.5}}$, ${CO_2}$ in the air. According to the International Energy Agency (IEA), the global energy demand decreased by 3.8\% in the first quarter of 2020 compared to the same period of 2019 because of the sudden reduction in economic activities and mobility\cite {international2020global}. Many recent studies on urban air quality have estimated the impacts of lockdown and confinement measures on various criteria pollutants. These studies mainly estimate the business as usual concentration of the pollutants for 2020 based on climate variables using ML algorithms. Finally, the impacts of the lockdown and confinement on air quality were assessed by comparing the estimated baseline concentration with actual concentration of pollutants in 2020. Some studies explored the relationship between pre-existing air pollution and COVID-19 mortality rates in different sections of cities.

Considering the magnitude of these effects, this study has been conceptualized as a survey of previous research on the impacts of the COVID-19 pandemic on urban mobility and air quality and their associations. However, this study focuses particularly on past studies that used various machine and deep learning approaches to evaluate the relationships among these concepts and conditions. With the advent of novel data technologies, methods of ML have been used extensively in disease prediction \cite {goldstein2017moving, zhai2020using, sujath2020machine}, transportation modeling \cite {angarita2020general, zhang2020nonlinear}, economic analysis \cite {barboza2017machine, abdella2020sustainability}, environmental modeling \cite {jang2020integrated, rahmati2020machine}, public sentiment analysis \cite {mishev2020evaluation, yadav2020sentiment}, etc. due to their remarkable computational ability to extract meaningful relationships between input and output features from large and complex datasets that are semantically diverse and that exhibit heterogeneous spatial and temporal granularity \cite {ben2020impact}. The problem at hand exhibits multiscalarity (from the individual, to the family, the community, the region, the nation, and eventually the global humanity), and other properties such as endogeneity, nonlinearity, non-independence, ambiguity and contextuality, that legitimately make them so-called "wicked problems" \cite {churchman1967guest, conklin2006dialogue}. ML has proved very effective to tackle such problems. Learning-based algorithms can retrieve meaningful features from a large volume of data to predict outcome accurately and able to reveal the hidden patterns in the data set that were previously unknown \cite {deshpande2020disease, nilashi2017212, senders2018478}. At variance with traditional data processing systems,  ML algorithms build models based on existing data with little or no distributional requirements for future predictions or decision making, which increases their performance tremendously \cite {shafique2020robust}.

In this regard, many studies have used ML and deep learning along with the SEIR (susceptible-exposed-infectious-recovered) model to predict COVID-19 transmission rates and evaluate its impacts on public health, urban mobility and the environment. Thus, considering the model prediction accuracy and the inherent power to explore big data, this study purports to review the literature that only used ML and deep learning based approaches. The main objective of this study is pursued by investigating the following research questions:

\begin{itemize}
    \item What are the impacts of the COVID-19 pandemic on mobility patterns of urban populations?
    \item What are the impacts of the COVID-19 pandemic on urban air quality?
    \item How do the different aspects of COVID-19 pandemic, human mobility, and air quality interact with each other?  
\end{itemize}

The salient contributions of this paper are five-fold: 
\begin{itemize}
    \item Understanding and quantifying the impacts of the COVID-19 pandemic on human mobility and air quality by reviewing past literature; 
    \item Identifying data sources and ML approaches that have been used in the previous studies and could be used by these researchers to estimate the impacts of COVID-19 on mobility reduction and on improving air quality in urban and rural areas;
    \item Developing a conceptual framework to clearly articulate the complex relationships among COVID-19 reported cases (and deaths), lockdown and confinement measures, human mobility patterns, and factors of air quality; 
    \item Identifying policy options that could be used by decision makers and researchers to reduce the severity of the pandemic, facilitate human mobility, and improve air quality; 
    \item Identifying potential research options that could be adopted by the researchers in their future work.
\end{itemize}

\section{Conceptual framework}
\begin{figure*}[htbp]
    \centering
    \includegraphics[width=\linewidth]{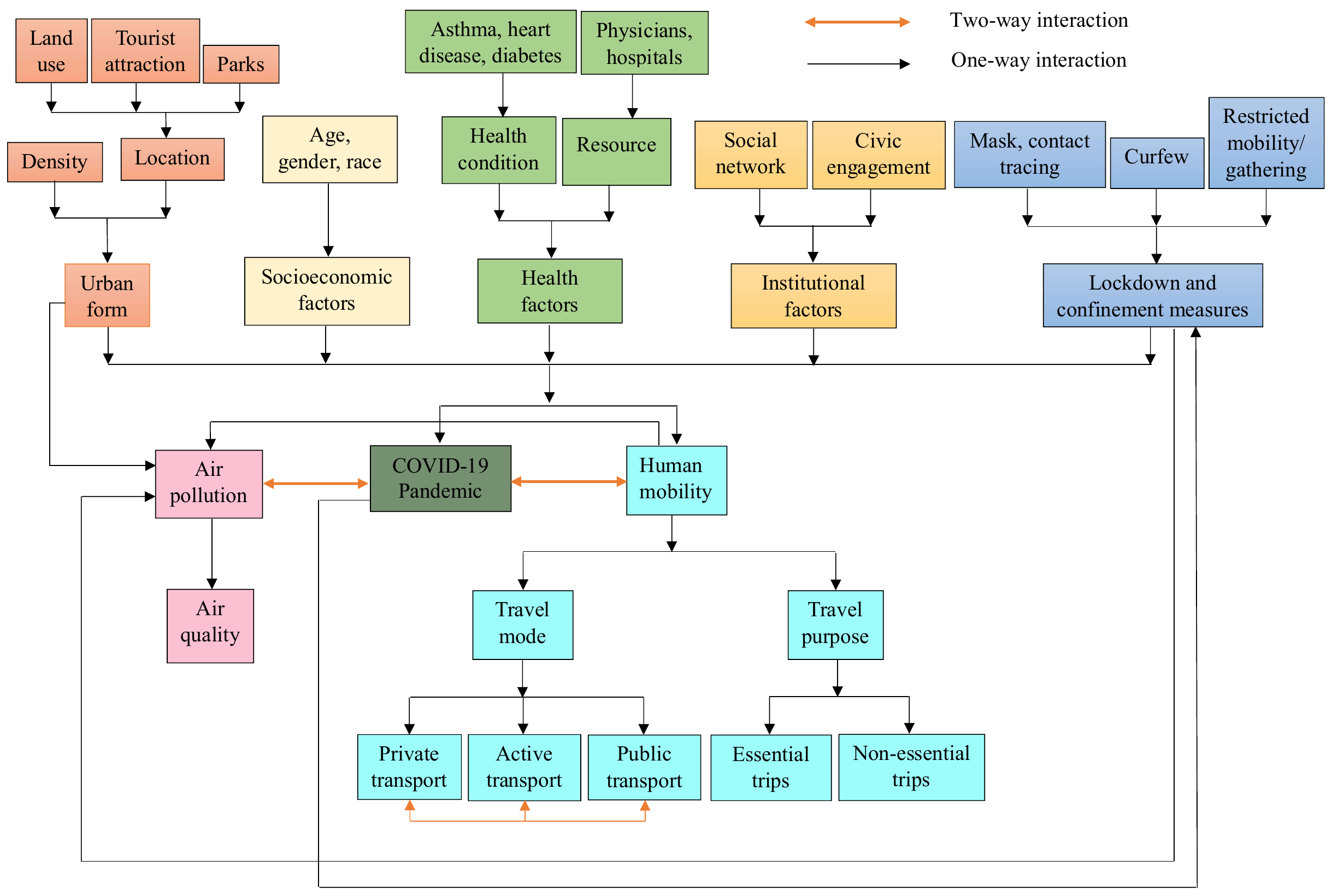}
    \caption{Conceptual framework of the study.}
    \label{Fig:concept}
\end{figure*}

Based on existing theories and concepts, a conceptual framework has been developed in this review study, Fig. \ref{Fig:concept}. The figure depicts that the urban form and structure, socioeconomic features of the people, health factors, social networks, civic engagement, and different lockdown and confinement measures significantly influence the COVID-19 pandemic. For example, high population density and public places are positively associated with COVID-19 because of increased interaction among the people. Elderly people and people with certain health conditions are more vulnerable to COVID-19. Similarly, limited access to health resources (e.g., hospitals, clinics, physicians) increases the risks of the pandemic. On the other hand, strong agency and institutional compliance (i.e., limited social engagement) and lockdown and confinement measures significantly reduce community transmission of COVID-19. 
These factors also influence human mobility patterns, travel mode choice behaviors, and travel purposes. For example, usually, the people in the high-density areas are more likely to use public transportation, active transportation (e.g., walk, cycle), and less likely to use private cars. But during the COVID-19 situation, the elderly, children, and disabled people are less likely to use public transport and more likely to use private cars to reduce infection risk. The lockdown and confinement measures significantly reduce human mobility in both essential (e.g., work, healthcare facilities) and non-essential (e.g. parks, fitness centers) trips. People adjust their travel schedules and change travel patterns to avoid infection risks. Thus, there is a bidirectional relationship between human mobility and the diffusion of the virus. The severity of the COVID-19 pandemic also influences governments and decision makers to impose lockdown measures to reduce the infection rate or slow the pace of growth (the so-called "flattening of the curve").   

Air quality significantly depends on urban form and structure. Industrial development, urbanization, and higher use of private transportation degrade air quality in urban areas. However, air quality has significantly improved during the COVID-19 lockdown periods thanks to reduced human mobility and the closure of many workplaces and industries. Improved air quality, in turn, lessens COVID-19 related deaths, particularly among people with chronic respiratory illnesses. Thus, similar to human mobility, COVID-19 has a bidirectional association with air quality.
    
\section{Human mobility patterns}
This section discusses the interrelationships among the COVID-19 transmission, confinement measures, socioeconomic factors,  human mobility and travel patterns of the people, etc., based on prior studies that used ML techniques. A summary of this literature is provided in Table \ref{tab:geoCOntext}.

\begin{table*}[]
\centering
\caption{Geographical context and objectives of past studies of COVID-19 and human mobility.}
\label{tab:geoCOntext}
\begin{tabular}{|p{1cm}|p{5.5cm}|p{9.5cm}|}
\hline
\textbf{Authors} & \textbf{Study context} & \textbf{Objective} \\ \hline \hline
\cite {roy2020characterizing} & Los Angeles, US & Investigating the relationships among socio-economic features of people and human mobility during COVID-19. \\ \hline
\cite {al2020measurement} & 13 countries of the world & Evaluating the effectiveness of lockdown measures on the COVID-19 pandemic. \\ \hline
\cite {wang2020agent} & New York and Seattle, US & Investigating the impacts of post-COVID-19 reopening strategies on travel patterns and mode choice of people. \\ \hline
\cite {szczepanek2020analysis} & Cracow, Poland & Investigating changes in pedestrian activities in public places (e.g., tourist spots, residential areas, and places with mixed land uses) before and during COVID-19. \\ \hline
\cite {sruthi2020predicting} & 50 states of the US & Assessing the impacts of policy instruments (e.g., closing and reopening of retail stores, workplaces, businesses, places of entertainment and worship, and restriction on mobility) on the COVID-19 pandemic. \\ \hline
\cite {soures2020sirnet} & US, Italy, Spain, Germany, France, and South Korea & Understand the impacts of social distancing measures (i.e., mobility) on the transmission of COVID-19. \\ \hline
\cite{scarpone2020multimethod} & 401 counties in Germany & Exploring the spatial (e.g., population density) and aspatial (e.g., socio-economic) factors of coronavirus diffusion \\ \hline

\cite {polyzos2020tourism} & US and Australia & Forecasting the effects of the COVID-19 pandemic on the tourists' arrivals. \\ \hline
\cite {chakraborty2020linear} & 50 states of the US and  District of Columbia & Investigating the factors that affect human mobility and travel during the COVID-19 pandemic. \\ \hline
\cite {delen2020no} & 26 countries of the world & Examining the role of social distancing measures on the COVID-19 transmission rate. \\ \hline
\cite {ayyoubzadeh2020predicting} & Iran & Predicting coronavirus cases and identifying the associated factors that influence new daily cases. \\ \hline
\cite {bao2020covid} & Boston,   US & Estimating the changes in people’s mobility due to COVID-19 situations and related local policy measures. \\ \hline
\cite {ahmed2020deep} & Hayatabad, Pakistan & Detecting the violation of social distancing measures. \\ \hline
\cite {chiang2020hawkes} & US & Modeling COVID-19 transmission at the county level. \\ \hline
\cite {yao2020impact} & Detroit, US & Investigating the impacts of COVID-19 and social distancing measures on traffic volume and safety. \\ \hline
\cite {hou2020intra} & Dane and Milwaukee County, City of Madison, US & Modeling of COVID-19 spread and investigating the associations between COVID-19 transmission and mobility, business foot-traffic, and socioeconomic features. \\ \hline
\cite {iwendi2020covid} & Wuhan, China & Predicting the number of COVID-19 infection cases related to patient recovery and death. \\ \hline
\cite {kuo2020evaluating} & 3219 counties in the US & Developing a COVID-19 case prediction model with ML techniques based on county-level data. \\ \hline
\cite {InteractiveCOVID19} & US & Developing an interactive platform to analyze COVID-19 impact. \\ \hline
\cite {Lu2020CovidForcast} & China & Predicting COVID-19 cases on the next day. \\ \hline
\cite {shirvani2020correlation} & Iran & Investigating the impacts of air and inter-city travel on COVID-19 confirmed cases. \\ \hline
\end{tabular}
\end{table*}

\subsection{Impacts of the COVID-19 pandemic on mobility and travel patterns}
\begin{figure*}[h]
    \centering
    \includegraphics[width=7in]{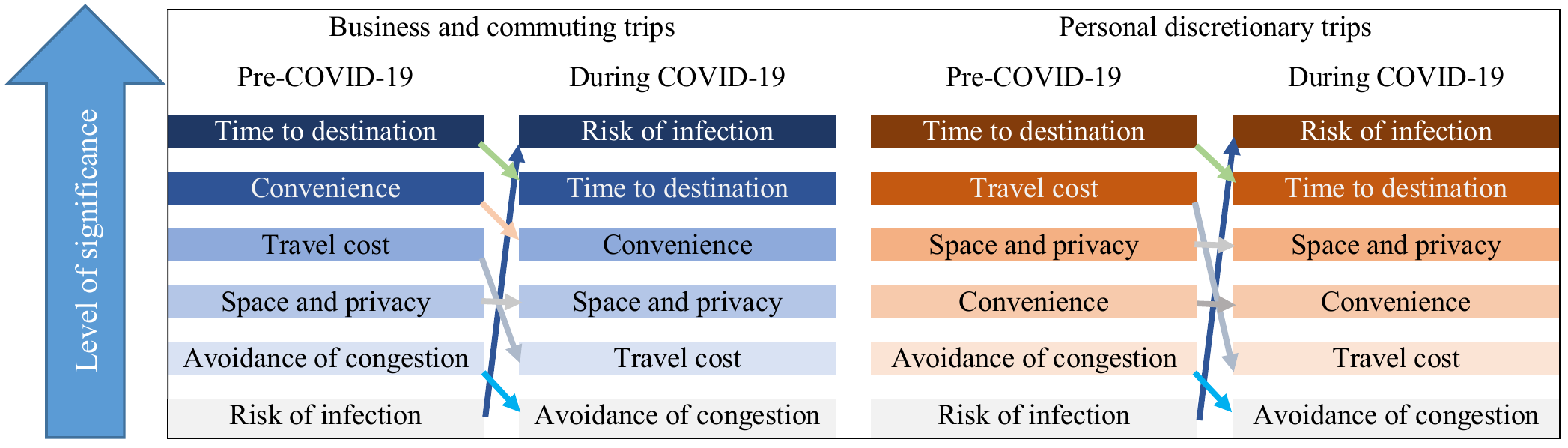}
    \caption{Key reasons to choose transportation modes before and during COVID-19 pandemic \cite{McKinsey2020five}.}
    \label{Fig:keyreasons}
\end{figure*}

It is observed that COVID-19 significantly influences people's travel habits for both essential and non-essential trips.
The McKinsey Center for Future Mobility (MCFM) \cite{McKinsey2020five} studies the key factors impacting travel mode choice behaviors of people in China, France, Germany, Italy, Japan, the UK, and the US before and during the COVID-19 pandemic, Fig. \ref{Fig:keyreasons}. Travel time, cost, and convenience played a significant role in people's travel mode choices for business, commuting, and personal trips before COVID-19 and similarly disruptive conditions. However, during this COVID-19 pandemic, reducing the risk of infection has become people's primary factor in deciding on their travel mode. Thus, the use of personal cars, cycling, walking, and shared micro-mobility is outpacing the use of various forms of public transportation.

Wang \textit{et al.} \cite {wang2020agent} investigated the impacts of post-COVID-19 reopening strategies on travel patterns and mode choices of the people in New York and Seattle, US. Upon collecting data, these authors observed a reduction in subway ridership in May 2020 compared to the 2019 level. An agent-based simulation demonstrates that full reopening could only expect about 73\% of the pre-pandemic transit ridership, but on the flip side, an increase in private car ridership by 142\%. Similarly, the number of walking trips and bike trips would increase by 101\% and 104\% of the pre-pandemic time, respectively. Applying a capacity restriction on public transit to 50\% for practicing social distancing, the study observed a decrease in transit ridership to 64\%, whereas car trips would increase by 143\% and bike trips by 123\%. Using a deep-learning based real-time video processing method of feature recognition, they found an increasing number of pedestrians at multiple locations in New York in May. They also observed a reduction in pedestrians who are practicing 6-feet social distancing guidelines from 91\% on April 2 to 86\% on May 27. Although pedestrian density was found low at some other locations, car and pedestrian density began to increase during the peak hour periods due to travel of essential workers, but the density of bicyclists was similar to the pre-pandemic situation. In a similar study context (i.e., Detroit, US), researchers in \cite {yao2020impact} investigated the impacts of COVID-19 pandemic and social distancing measures on traffic volume on the transportation network and on safety. Collecting data from 73 signalized intersections, they mentioned that, aside from the continued dominance of car trips, the number of bicycle and motorcycle trips has mushroomed four-fold during the COVID-19 pandemic. The number of trucks and vans remains unchanged before and during the confinement and lockdown periods. However, a 40\% increase in the number of trucks and vans was observed when restrictions were lifted. Thus, the COVID-19 pandemic, reopening strategies, and confinement and lockdown measures have severe impacts on personal mobility and on the state of the urban transportation system more generally.

In a recent survey, MCFM \cite {McKinsey2020survey} discovered that about 70\% of respondents would choose walking or cycling at least once per week, even after returning to normal life, which is more than 6\% higher than the pre-crisis situation. Similarly, private car-sharing would increase from 78\% to 79\% in the post-pandemic scenario. However, despite a dramatic drop during the pandemic, transit ridership would return to its pre-existing conditions at around 40\% on average across all surveyed countries. Similarly, shared micro-mobility and car-sharing would be slightly more popular (i.e., 1 to 2\%) after returning to normal life. Thus, while travel mode choice behaviors of the people are found to be significantly affected by the COVID-19 pandemic, people's overall travel patterns would be expected to bounce back to their state before the pandemic.

COVID-19 has also significantly affected the social and economic aspects of tourism and associated industries (e.g., air transportation) in major tourist destinations around the world. Polyzos \textit{et al.} \cite{polyzos2020tourism} forecasted the effects of the COVID-19 pandemic on Chinese tourist arrivals in the US and Australia. Using data from the 2003 SARS outbreak, they trained a deep-learning-based Long Short-Term Memory (LSTM) neural network model to predict the social and economic impacts of the pandemic on the tourism industry. The trained model is calibrated considering the particulars of current situations (e.g., lockdown, flight bans) to simulate the impacts of a COVID-19-like pandemic on the tourism industry. The results indicate a significant drop in tourist arrivals from China to the US and Australia. Upon cross-validating the findings based on backtesting techniques (i.e., the sample is split into smaller training sets and error is used to train the model), the researchers commented that it would require 6 and 12 months for tourist arrival rates to Australia and the US, respectively, to recover to their pre-pandemic levels after the recent collapse of the industry. They also mentioned that the LSTM technique performs better than other artificial neural networks (e.g., hidden Markov, Support Vector Regression) and forecasting models, such as ARIMA (Auto-Regressive Integrated Moving Average), to predict the impacts of COVID-19.

The changes in pedestrian activities in public places (e.g., tourist attractions, residential areas) because of COVID-19 have been investigated in \cite{szczepanek2020analysis} for Cracow, Poland, using YOLO (You Only Look Once) - an ML algorithm that allows easy and less erroneous end-to-end object detection. After collecting data from webcams located at public sites covering the period of June 9, 2016, to April 19, 2020, the images are first split into smaller tiles, which increases pedestrian detection capacity by more than 50\%. Estimating hourly, daily and weekly averages of pedestrian activities, the study observed a 34-50\% reduction in pedestrians in residential zones and a 78-85\% reduction in tourist localities due to lockdown and confinement measures during the Covid-19 pandemic. The study claimed that the proposed method is more efficient to detect and count pedestrians from time-lapse webcam images than other approaches such as Single Shot MultiBox Detector (SSD), existing YOLO and Faster Region-based Convolutional Neural Network (R-CNN), in terms of mean absolute error (i.e., 4.28 vs 9.87, 5.48, and 5.38, respectively) and root mean square error (i.e., 7.96 vs 14.32, 10.23, and 9.16, respectively). However, it has a longer processing time (3.28s) compared to existing YOLO (0.75s) and SSD (1.46s), although a lower processing time than Faster R-CNN (8.35s).

Changes in human mobility and activity patterns can also be evidenced through the impact the pandemic would have in other economic sectors such as energy consumption. The pandemic oil demand analysis (PODA) is a ML technique proposed in \cite{Ou2020Gasoline} to project the US gasoline demand during the COVID-19 pandemic. It consists of two projection modules. The first, dubbed the Mobility Dynamic Index Forecast Module, identifies the changes in travel mobility caused by the evolution of the COVID-19 pandemic. The second, the Motor Gasoline Demand Estimation Module, estimates vehicle miles traveled on pandemic days, while considering the dynamic indices of travel mobility, and quantifies motor gasoline demand by coupling the gasoline demands and vehicle miles traveled. The prediction model used Apple and Google mobility data and showed a significant reduction in US gasoline demand in March and early April of 2020. Another study \cite{chen2020using} proposed to use mobility data as a complementary component in a day-ahead electrical load forecast model based on a multi-task neural network and showed that the load forecasting accuracy can be improved significantly. This study also found that the sudden changes in electric consumption due to COVID-19 pandemic cause higher forecasting error.

The above discussion clearly demonstrates that COVID-19 has significant impacts on urban transportation systems by influencing people's travel and mode choice behaviors and their energy consumption. During this pandemic, people adjust not only the sheer volume of their travel, but also their travel schedule, route, and modes to reduce potential health risks to themselves and to others. Consequently, it is also indirectly affecting other sectors of the economy (e.g., office businesses, manufacturing, retail and services). Overall, this pandemic is affecting human lives and the economic system of society very broadly and very deeply.     

\subsection{COVID-19 prediction models to understand the factors affecting virus diffusion}
Many researchers have used Artificial Intelligence (AI) to predict coronavirus infection rate, recovery, and death rate with good accuracy throughout the world. These researchers used compartmental models (such as the Susceptible-Exposed-Infectious-Recovered (SEIR) model), ML based models, and hybrid models to predict the pandemic itself. For example, the YYG model \cite{YYG2020Projections} uses a simple ML technique to estimate the number of COVID-19 infection cases and deaths in a state/country. This projection model adopted the SEIR model as the underlying simulator for generating simulated infections under specific scenarios and the simple brute-force grid search technique as the ML model for tuning the model parameters. The model used two types of parameters: fixed parameters (e.g., latency period, infectious period, time to recovery) and dynamic parameters (e.g., reproduction rate, mortality rate, mitigation effects). 


Similarly, many studies have investigated the factors that affect the COVID-19 transmission rate in cities and regions on the basis of an assumed COVID-19 transmission model. For example, Yao \textit{et al.} \cite {yao2020impact} evaluated the role of transportation (e.g., traffic volume, accident), social distancing measures, and weather conditions in the incidence of coronavirus confirmed cases. Calculating correlation coefficients, they found that daily confirmed cases are highly correlated with the transportation volume (e.g., cars), total crashes, social distancing indicators, and average temperature. Using the LSTM approach, they estimated the number of confirmed cases per day. Model results demonstrate that the inclusion of all selected features improved the performance of the model to predict daily confirmed cases with lower RMSE (0.0606), mean absolute error (MAE) (0.0378),
and a high \(R^2\) (0.9088). Thus, besides daily confirmed cases and the social distancing indicator in the previous days, the inclusion of traffic volume, crashes, and weather conditions significantly improves the predictive performance of the model. Considering daily inter-state traffic and air traffic data, including the number of transfer passengers, Shirvani \textit{et al.} \cite{shirvani2020correlation} predicted new confirmed cases in Iran. Using a supervised ML model consisting of an ensemble of linear regression, LASSO regression, K-Nearest Neighbor (KNN) regression, random forests, the model predicted the number of new cases with an accuracy of 85\%. The study found a positive correlation between inter-state travel and new confirmed cases. Consequently, it suggested imposing travel restrictions to limit COVID-19 transmission and slow the spread of the pandemic.

In the US, Kuo and Fu \cite{kuo2020evaluating} developed a COVID-19 prediction model after collecting demographic, environmental, and mobility data at the county level. Data from 172 metropolitan counties were used to design a hybrid framework based on eight different ML algorithms to predict daily and cumulative confirmed cases. The final model was developed using a general linear model (GLM) that combines the predictions from all ML algorithms. The study showed that human mobility in the metropolitan areas reduced substantially after implementing lockdown measures in mid-March. Scenario assessment results show that a 1- week and 2-week lockdown in Phase I reopening could reduce infections by 4-29\% and 15-55\% in the future week, respectively. However, a 2-week reopening in Phase II could increase infections by 16-80\%. Thus, this study suggested a mandatory lockdown order lasting more than one week to control the COVID-19 pandemic by reducing community mobility and transmission. Researchers in \cite{luo2020distribution} investigated the non-linear relationships of COVID-19 death rates with environmental, health, socioeconomic, and demographic risk factors using geographically weighted random forests (GW-RF) non-parametric regression model in the US. Collecting county-level daily death counts in 3108 counties, the model showed that walking trips to work, concentration of air pollutants, households with a mortgage, unemployment status, and percent of black or African Americans have a strong correlation with the spatial distribution of COVID-19 incidence with an \(R^2\) of 0.78.

Another study in the US \cite{hou2020intra} developed a stochastic SEIR-style epidemic model augmented by human mobility to predict historical growth trajectories of COVID-19 cases in two counties (Dane and Milwaukee) in the state of Wisconsin. The model was combined with the data assimilation (i.e., Kalman Filter) and ML techniques (i.e., Walktrap network) to reconstruct the COVID-19 trajectories. The Walktrap network partitions the counties into clusters based on the observed human mobility data. A local SEIR model for each region was developed using geographic, socioeconomic, cultural and transportation factors of the people to get a region-specific effective reproduction number. Finally, the combined model helps to investigate the associations of COVID-19 diffusion with mobility patterns, business foot-traffic, race, and age groups. The study found a strong association between reproduction number and visits to drinking establishments (alcoholic beverages). It is suggested that policy makers should explicitly consider the local transmission scenario even after restricting intra-regional movement for preventing more health disparities in future pandemics. In another North American country (i.e., Labrador and Newfoundland in Canada), Kevln \textit{et al.} \cite{Linka2020Reopening} explored the impact of partial and total reopening of airports on COVID-19 outbreaks with SEIR and Bayesian Interference models. Using data on air-traffic mobility incoming and outgoing from/to other Canadian provinces and the US with different measures of quarantine, the study found that relaxing travel restrictions is possible entirely (total reopening). However, strict (100\%) quarantine conditions are necessary to control the disease outbreak. Voluntary quarantine, even at an overall rate of 95\%, is not enough to entirely prevent future outbreaks. Thus, strict policies on quarantine after entering the city are essential to control the massive outbreak of the pandemic.

In China, Lu \textit{et al.} \cite{Lu2020CovidForcast} developed an ensemble-based back-propagation neural network (BPNN) model for predicting COVID-19 cases on the next-day. Using Baidu‘s migration data (e.g., migration index, internal travel flow index and, confirmed cases from the 13 previous days), they trained the model for predicting the coronavirus cases and achieved a 97\% correlation with the actual data. Multiplying Baidu’s mobility index values by two and considering the incremented values due to mild government interventions on human mobility, this study reported a significant increase in the COVID-19 cases and continuous growth for a long time. Thus, mobility has significant impacts on COVID-19 transmission. However, the researchers mentioned that the inclusion of other social distancing measures could increase the accuracy of the prediction. Another study in Wuhan, China \cite{iwendi2020covid}, proposed a fine-tuned Random Forest model boosted by the AdaBoost algorithm for predicting the coronavirus cases and their possible outcomes (i.e., recovery, death). Collecting information on COVID-19 patient’s geographic location, travel, health, and demographic features, the study developed the prediction model using a grid search algorithm which provides a set of best-performing parameters. The model results indicate that the proposed fine-tuned model correctly predicts the outcomes with an accuracy of 94\% and an F1 score of 0.86. The study found higher death rates among the Wuhan natives compared to non-natives. The study also reported higher death rates of male patients than female patients and the majority of the patients were from 20 to 70 years old.

\begin{figure}[ht]
    \centering
    \includegraphics[width=3.6in]{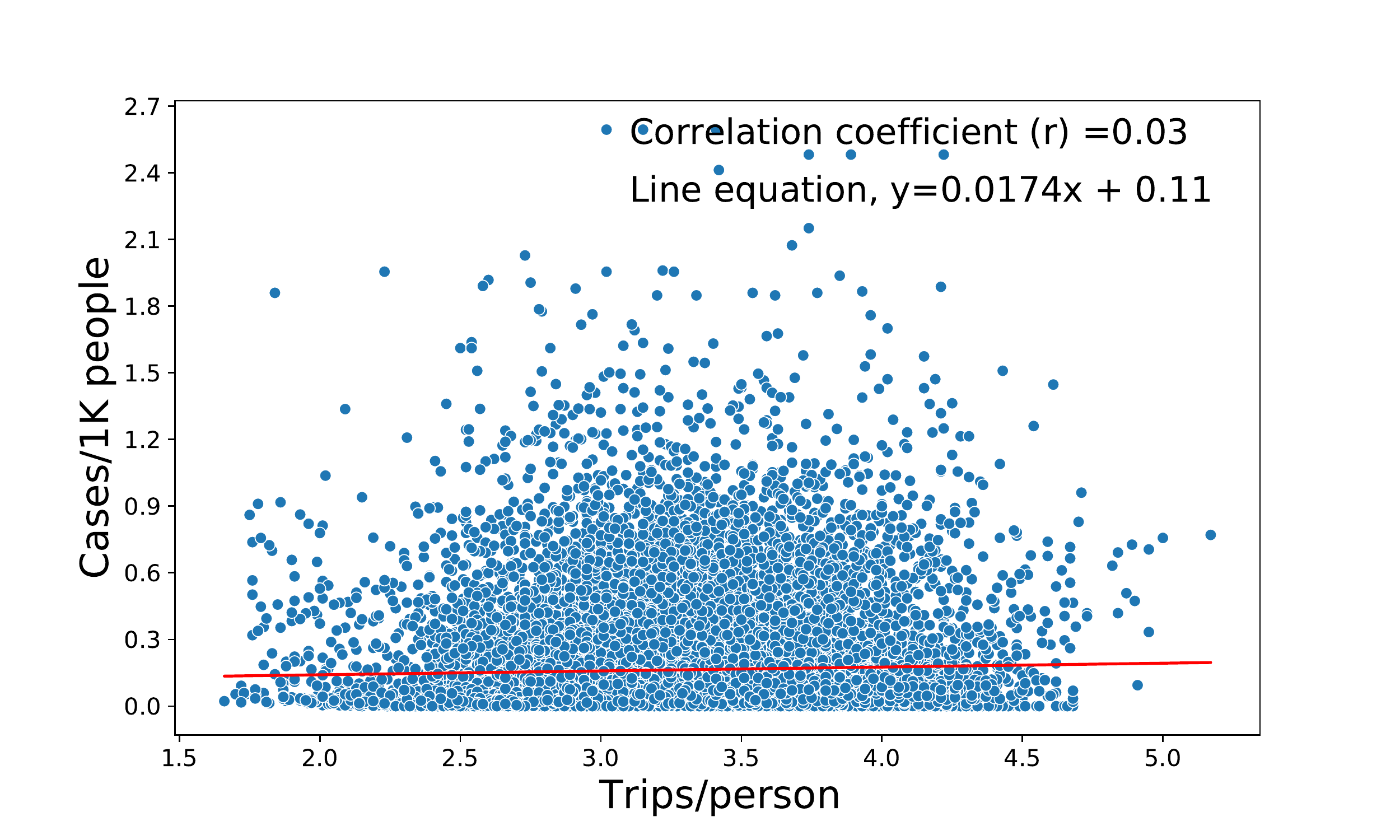}
    \caption{Impact of mobility on COVID-19 infection.}
    \label{Fig:trip_case}
\end{figure}

In summary, ML is widely used in predicting COVID-19 confirmed cases, deaths, and the recovery status of populations (Table \ref{tab:mobilityDataML}). These models use a variety of transportation, demographic and socioeconomic, geographical, environmental, health, and lockdown-related factors to predict confirmed cases and deaths. The results from these models indicate that ML can efficiently predict the COVID-19 pandemic with high rates of accuracy. Most of the studies suggested putting an embargo on the movement of people to prevent a widespread pandemic. Yet, some the evidence that has accumulated points in other directions. For instance, with data on daily trips per person and COVID-19 confirmed cases in different US states, \cite {COVID19Platform} found a positive but weak association between the number of trips per day and the incidence of COVID-19 confirmed cases, Fig. \ref{Fig:trip_case}. More analytic research that comprehensively controls for a full range of factors and circumstances is needed to have a robust knowledge base in which to ground policy making.  

\begin{table*}[]
\centering
\caption{Data and ML methods used in past studies on COVID-19 and human mobility.}
\label{tab:mobilityDataML}
\begin{tabular}{|p{1cm}|p{9.5cm}|p{5.5cm}|}
\hline
\textbf{Author} & \textbf{Data and sources} & \textbf{ML methods} \\ \hline \hline
\cite {roy2020characterizing} & Mobility to different points of interest (POI) such as restaurants, grocery stores, religious establishments, fitness centers, and supermarkets from SafeGraph, social vulnerability index (i.e., socioeconomic status, household composition and disability status, minority status and language proficiency, housing type and transportation) from CDC, unemployment rate from American Community Survey, population density from WorldPop, and $NO_2$ concentration from NOAA, Daily COVID-19 cases from LA Public Health Department & Gradient Boosting Decision Tree (GBDT) \\ \hline
\cite {al2020measurement} & Mobility patterns at retail and recreation, grocery stores and pharmacies, parks, transit stations, workplaces, residential areas, and schools from Google Mobility Report and UNESCO & Random Forest (RF), K-Nearest Neighbor (KNN) \\ \hline
\cite {wang2020agent} & Subway ridership and vehicular traffic data from the Metropolitan Transportation Authority (MTA), transit, driving, and walking trips from Apple Mobility Trends reports & Agent-based simulation (MATSim-NYC), deep-learning-based real-time video processing method (e.g., RetinaNet, ResNet-50) \\ \hline
\cite {szczepanek2020analysis} & Time-lapse webcam images from different urban zones (e.g., tourist areas, residential areas) for detecting pedestrian and their activities & YOLO, Single Shot MultiBox Detector (SSD), and Faster R-CNN \\ \hline
\cite {sruthi2020predicting} & Information on policy instruments from the press release from states and mobility data from google mobility reports & Extreme Gradient Boosting Decision Tree (XGBDT) \\ \hline
\cite {soures2020sirnet} & Coronavirus case count from WHO, CDC, NY Times, and Texas DSHS, socioeconomic data from US census, data on mobility changes in retail and recreation, grocery and pharmacy, parks, transit stations, workplaces, residential from Google and Apple & SIRNET - a hybrid model comprising of Recurrent Neural network (RNN) and SEIR model \\ \hline
\cite{scarpone2020multimethod} & COVID-19 cases from the county, socioeconomic data (e.g., age, gender, education, income) from various ministries, federal states, and municipal governments, built environment (e.g., density of airports, train stations, grocery stores, parks) from OpenStreetMap & Bayesian Additive Regression Trees (BART) model \\ \hline

\cite {polyzos2020tourism} & Monthly   Chinese tourist arrival data from the 2003 SARS outbreak up to October 2019 from the National Travel \& Tourism Office and Australian Bureau of Statistics & Long Short-Term Memory (LSTM) model \\ \hline
\cite {chakraborty2020linear} & COVID-19 cases and deaths, number of trips, social distancing index, out of county trips, transit mode share, population density, population staying at home, working from home, socioeconomic information (e.g., race, income, and employment status, unemployment rate) from University of Maryland (UMD), age category, education, male to female ratio, gross domestic product from US Census Bureau & Ridge, LASSO, Elastic Net modeling techniques \\ \hline
\cite {delen2020no} & COVID-19 data from the European Centre for Disease Prevention, mobility in retail and recreation, grocery and pharmacy, parks, transit stations, workplaces, and residential areas from Google and driving, walking, and transit trips from Apple & Gradient Boosting Decision Tree (GBDT) \\ \hline
\cite {ayyoubzadeh2020predicting} & Coronavirus daily incidence from Worldometer, number of searches for Antiseptic, Handwashing, Hand sanitizer, and Ethanol from google search trend & LSTM \\ \hline
\cite {bao2020covid} & COVID-19 cases and deaths from the CDC, POI visits data from SafeGraph, population and income from US census, policies such as the closure of public venues and schools from local government & COVID-GAN (Conditional generative neural network approach) \\ \hline
\cite {ahmed2020deep} & An indoor video data recorded at the Institute of Management Sciences, Hayatabad, Peshawar Pakistan, which contains video sequences captured from the   overhead view & You Only Look Once (YOLO) \\ \hline
\cite {chiang2020hawkes} & COVID-19 cases and deaths from New York Times, mobility index from Google mobility reports, socioeconomic data (e.g., population, age, number of hospitals and  ICU beds, percentage of smokers and diabetes, and heart disease mortality) from CDC & Hawkes process using the expectation-maximization (EM) algorithm \\ \hline
\cite {yao2020impact} & Transportation data (e.g., traffic volume, accident) from 73 signalized intersections, COVID-19 cases from Michigan’s official Coronavirus dashboard, weather factors (e.g., temperature and wind speed) from National Oceanic and Atmospheric Administration, social distancing data from UMD, Crash Data from Michigan State Police (MSP) & Long Short Term Memory (LSTM) \\ \hline
\cite {hou2020intra} & COVID-19 cases from local COVID-19 Dashboards, census-tract geographic boundaries and socioeconomic attributes from US Census Bureau, POIs with human travel patterns from SafeGraph & Walktrap network-based community clustering method, modified SEIR model, Ensemble Kalman Filter \\ \hline
\cite {iwendi2020covid} & Coronavirus cases, deaths, patients age, gender from the World Health Organization and John Hopkins University & Decision Tree (DT) Classifier, Support Vector Machine (SVM) Classifier, Gaussian Naïve Bayes Classifier, Boosted RF Classifier \\ \hline
\cite {kuo2020evaluating} & Coronavirus cases and stay-at-home order from New York Times, socioeconomic (e.g., population density, labor force rate, unemployment rate, income) from US Department of Agriculture, Economic Research Service, weather information (e.g., temperature, wind speed, precipitation, solar radiation) from gridMET, mobility data from Google mobility reports & Elastic net (EN) model, KNN, DT, RF, GBDT, and Artificial Neural Network (ANN) \\ \hline
\cite {InteractiveCOVID19} & Mobile device location data based on GPS, call detail record, Cellular Network data, Social media Location-based Services, COVID-19 data from John Hopkins University, socioeconomic data from American Community Survey, travel information from National Household Travel Survey & Deep learning algorithms \\ \hline
\cite {Lu2020CovidForcast} & Migration data (e.g., migration index, \% of migrants coming from other cities, and internal travel flow index) from Baidu Maps LBS open platform, confirmed cases & Ensemble-based back propagation neural network (BPNN) model \\ \hline
\cite {shirvani2020correlation} & Car traffic data from the road management center of Iran, air flight traffic data from Iranian's airports and air navigation office, and COVID-19 daily new cases from John Hopkins University & LASSO   regression, KNN, RF, and ensemble learning \\ \hline
\end{tabular}
\end{table*}

\subsection{Impact of lockdown and confinement measures on mobility and COVID-19 pandemic}
Many studies have assessed the effectiveness of lockdown and confinement measures and travel restrictions imposed by state and governmental agencies to control the massive outbreak of the COVID-19 pandemic.
Al Zobbi \textit{et al.} \cite {al2020measurement} evaluated the effectiveness of these measures using data analytics and ML. Collecting data from Google mobility reports and UNESCO’s website, they estimated the mobility pattern of the people at retail and recreation sites, grocery and pharmacy outlets, parks, transit stations, workplaces, residential buildings, and schools during the period when the pandemic prompted lockdown orders and confinement directives. They used the reproduction number (R0) to represent the COVID-19 pandemic in 13 countries around the world. The daily R0 values were grouped into interquartile ranges and close monitoring of mean values was performed to check the efficiency of daily social distancing measures. Random Forest (RF) and KNN methods were used to establish the direct correlation between lockdown and confinement measures and pandemic severity. The results show a higher correlation (0.68) and the least MAE. Thus, these measures have a significant influence on the coronavirus infection rate. Dividing lockdown efficiency into four categories from Group A (highest efficiency) to B, C, and D (lowest efficiency), they found higher efficiency in Australia and South Korea. Efficiency improved significantly from group D to group B, group D to group C, and group C to B in Germany, Spain, and India, respectively. Despite not having strict lockdown measures, South Korea showed high efficiency due to widespread self-quarantine and self-awareness in the population. However, the US and Brazil showed lower levels of efficiency due to the late implementation of lockdown measures. The study concluded that the coronavirus infection rate dramatically dropped due to strict lockdown procedures, self-quarantine, and people’s awareness about the disease. 


By building a predictive model, researchers in \cite {sruthi2020predicting} assessed the impacts of different policy instruments (e.g., graded closing and reopening of retail stores, workplaces, businesses, places of entertainment and worship, and restriction on mobility) on the COVID-19 transmission rate in the 50 US states. With data on policy instruments collected from press releases and with data from Google's mobility reports for the period of March 09 to August 2 2020, they used XGBboost to predict the transmission of COVID-19 under different policy scenarios. With an \(R^2\) value of 0.79 for training and 0.76 for testing, the model showed a robust estimation of the transmission rate as a function of policy instruments. 
Thus, various policy instruments have a significant influence on future transmission. This study concluded that state agencies could ensemble policy instruments in a structured way for data-driven decision making. Similarly, Badr \textit{et al.} \cite{AssociationBadr2020} studies the social distancing index based on population mobility of 25 American counties. Using the mobility data from Teralytics (Zürich, Switzerland), they found a 35-63\% reduction in mobility compared to normal conditions due to  restrictions imposed on mobility. The study also shows a strong correlation between popuklation adherence to strict social distancing directives and COVID-19 cases reduction in the US. 

Soures \textit{et al.} \cite{soures2020sirnet} proposed a new hybrid ML model comprised of neural network and epidemiology models (SIRNET) to understand the impact of social distancing (i.e., mobility) on the spread of COVID-19 infections in the US, Italy, Spain, Germany, France, and South Korea. With data assembled from multiple sources (e.g., WHO, CDC, NY Times, US Census, Google, and Apple), they evaluated COVID-19 situations in different geographic regions. The results demonstrate that the SIRNET model is able to predict coronavirus cases by region using mobility information. They observed that low mobility in a population group has a significant impact on reducing the COVID-19 cases in this population. Thus, they suggested keeping mobility at least below 50\% of the nominal mobility in the immediate future to control the pandemic until herd immunity is achieved.

Researchers in \cite {delen2020no} examined the role of social distancing measures on COVID-19 transmission rates by integrating mobility data from Google and Apple, and COVID-19 data from the European Centre for Disease Prevention and Control in 26 countries of the world. The transmission rate was calculated using the susceptible-infected-recovered (SIR) model.
 Their Gradient boosted decision tree regression analysis indicated that mobility changes in retail and recreation businesses, grocery stores and pharmacies, parks, transit stations, workplaces, and residential areas due to social distancing policies explain about 47\% variation in the disease outbreak. Thus, controlling restrictions on people’s attendance and mobility in public places with high density of people are effective public health policy measures to mitigate the impacts of the pandemic.

\begin{figure}[htbp]
    \centering
    \subfloat[Impacts of social distancing on daily trips per person.]{
    \includegraphics[width=3.5in]{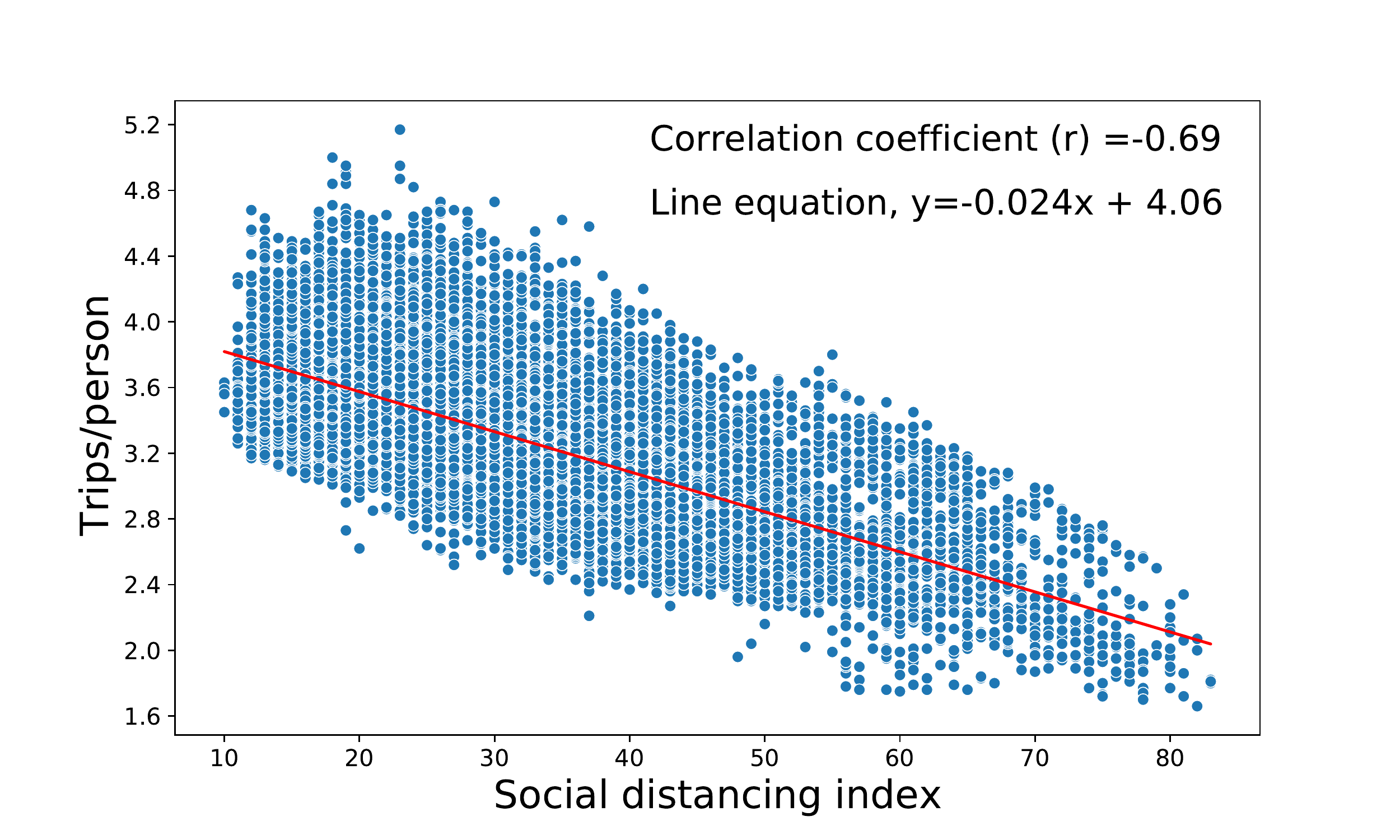}}
   
    \subfloat[Impacts of social distancing on COVID-19 cases.]{
    \includegraphics[width=3.5in]{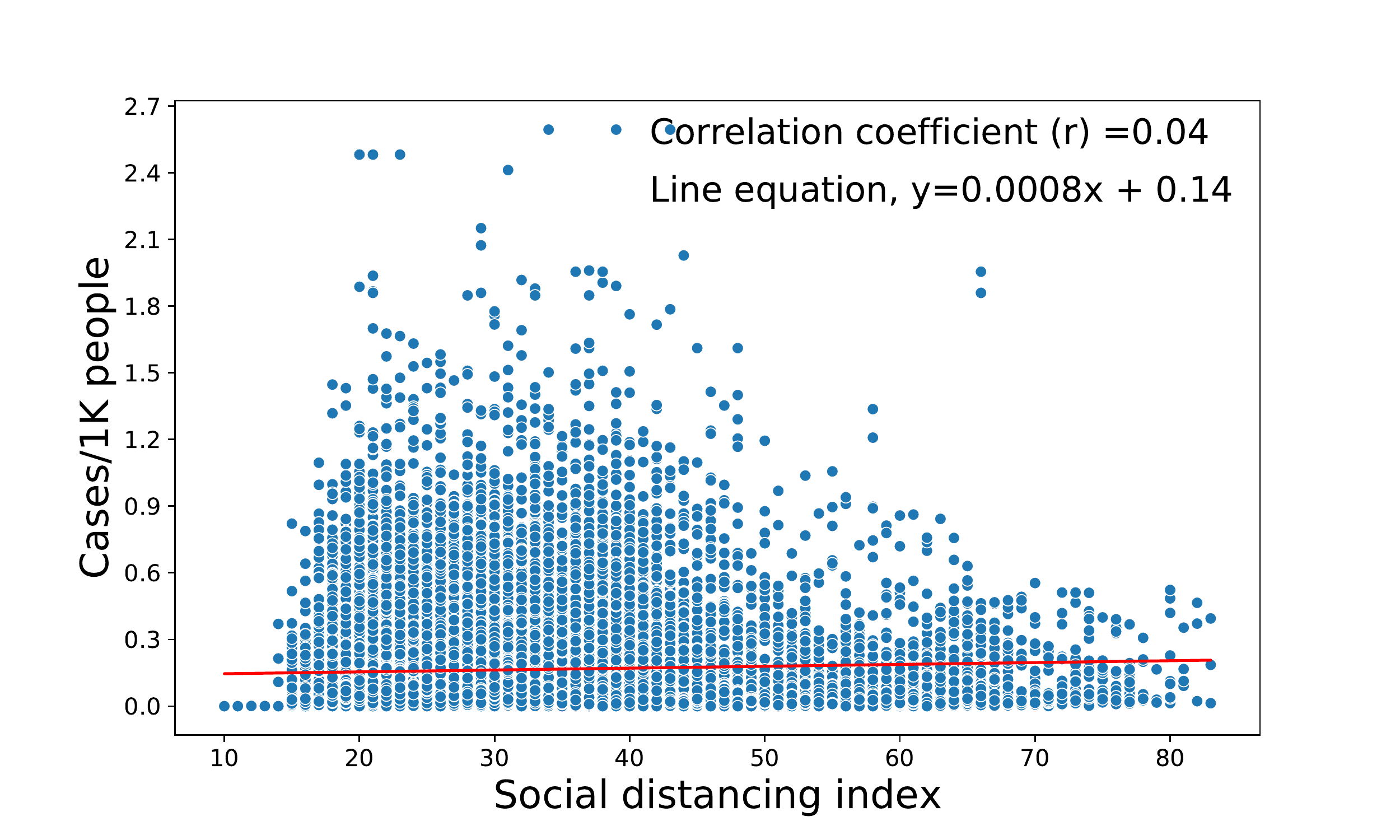}}
    \caption{Impacts of social distancing on mobility and COVID-19 pandemic.}
    \label{Fig:sdi_trip-case}
\end{figure}

It is now well documented that strict lockdown and confinement measures and proper practice of social distancing are very effective in mitigating the COVID-19 pandemic by reducing mobility and gathering of people. However, evidence much be used with great care and caution. To demonstrate this point, we further investigated the impacts of social distancing practices on human mobility and COVID-19 daily confirmed cases using the data collected from \cite {COVID19Platform} for the period of January 01 to December 26, 2020, Fig. \ref{Fig:sdi_trip-case}.
As anticipated, the results indicate a negative correlation of the social distancing index with daily trips per person. Intuitively, there should be a negative correlation between the incidence of COVID-19 cases and the daily trips per person; however, at first blush there is no significant correlation to be found. This emphasises the criticality of reliable data collected over time across varied communities, but also the importance of a modeling framework suitable for the handling of complex, multiple and possibly non-linear relationships and of the socio-spatial contexts of  individual decisions, social group dynamics, and public policy elements \cite {thill2020research}. Thus, we contend that further research using ML is fully warranted to reach strong conclusive statements.  


\subsection{Socio-economic factors affecting mobility and travel patterns during COVID-19 pandemic}
Besides the severity of the pandemic and related confinement and lockdown measures, other social, economic, and environmental factors may also affect the mobility and travel patterns of the population. A study in  \cite {chakraborty2020linear} investigated the factors that affect human mobility and travel in the US during the COVID-19 pandemic using OLS regression, ridge regression, LASSO, and Elastic Net modeling techniques. With statewide data from January 1, 2020, to June 13, 2020, and dividing them into train and test data sets, the study examined that ridge regression provides superior results with the least error. However, LASSO and Elastic Net modeling techniques performed better than OLS regression. The results indicate that the number of daily trips per person has a negative association with the number of new cases, social distancing index, median income, percentage of elderly population, number of people staying at home, socioeconomic status, stay-at-home order, and domestic travel restrictions. In contrast, the number of daily trips per person has a positive association with transit mode share, percentage of Hispanic and African American population, mandatory statewide mask policy, and driving mobility index. Finally, using LASSO regression, the study found that the percentage of the population over 60, social distancing index, and percentage of the population working from home have higher impacts on the number of daily trips per person compared to other covariates. 

Another study investigated spatial (e.g., population density) and aspatial (e.g., socioeconomic) factors of coronavirus transmission over 401 counties in Germany \cite{scarpone2020multimethod}. The Bayesian Additive Regression Trees (BART) model demonstrated that higher densities of churches were the most important factors for predicting COVID-19 cases due to higher walkability, interpersonal interactions, higher social connectivity, and community engagement, particularly among senior and elderly populations. Similarly, long-distance train stations increase the probability of inter-personal coronavirus transmissions due to the high density of travelers and their long-distance interactions. Among the socioeconomic factors, in-person participation in vote casting was identified as the strongest predictor of COVID-19 transmission. The study also noticed that the foreign guests in tourist establishments, employment density, community centers, beauty salons, etc., significantly impacted on COVID-19 transmissions. Therefore, they suggested implementing social distancing measures and reduce unnecessary travel to reduce coronavirus infection.

Researchers in \cite{roy2020characterizing} investigated the relationships between socio-economic characteristics of the people and human mobility in the context of the COVID-19 pandemic after collecting geolocated human mobility data from SafeGraph in Los Angeles, US. Quantifying mobility indices and social distancing metrics, they classified census blocks into areas of High, Medium, and Low vulnerability to the COVID-19 pandemic using supervised ML classifiers such as Gradient Boosting, Support Vector Machines, and Multinomial Logit models. The results indicated that tree-based classifiers (i.e., Gradient Boosting) performed well with an accuracy of 97.4\% and area under curve (AUC) score of 98.7\% compared to Random Forest (96.8\%), Support Vector Machines (90.5\%), and Multinomial Logit models (91.3\%). In substance, the study reported that socially vulnerable populations, high mobility indices and low social distancing index increases the vulnerability of the local communities to COVID-19 infection.

Using the expectation-maximization (EM) algorithm, the study in \cite {chiang2020hawkes} reported that the Hawkes process --a model that simultaneously estimates the intensity of the events and tracks dynamics of the reproduction number of the virus--has a good potential to predict COVID-19 transmission with minimum MAE and percentage error (PE). It also demonstrated that the performance of the prediction model increases substantially when the modeling of disease transmission is integrated with mobility, social distancing measures, and other socio-demographic covariates (e.g., population, median age, number of hospitals and  ICU beds, percentage of smokers and diabetic patients, and heart disease mortality). Estimating Poisson regression, the study mentioned that the reproduction number is positively associated with mobility changes in retail/recreation and grocery/pharmacy. In contrast, it is negatively associated with mobility changes in transit stations, parks, and residential areas. They also found that the reproduction number is positively associated with higher population density, number of hospitals, and ICU beds. In contrast, the reproduction number is negatively associated with median age, percentage of the population with diabetes, heart disease, and smoking habits, which imply that people with high health risks are more cautious and tend to live in areas with fewer coronavirus cases and population density, which protects them from getting infected.

The above discussion reveals the facts that in addition to COVID-19 related factors, socioeconomic situations of the people influence human mobility. For example, disadvantaged segments of society (e.g., elderly, people with heart disease) voluntarily limit their movement due to their concern and fear of exposure and infection. On the other hand, people who live in areas with more health care facilities would be more comfortable to travel around due to their preparedness to control any unwanted situations. Thus, socioeconomic, health and environmental factors have a significant impact on mobility and the COVID-19 pandemic.

\subsection{COVID-19 interactive platforms and dashboards}
A number of platforms have been developed to provide real-time COVID-19 related information such as new cases, deaths, testing, hospitalization, contact tracing, COVID-19 prediction, lockdown and confinement measures, impacts of COVID-19 on daily mobility, economy, and socioeconomic characteristics of people, etc. For example, Zhang \textit{et al.} provided a COVID-19 impact analysis platform to understand the daily impact of COVID-19 on mobility, economy, and society \cite{COVID19Platform,InteractiveCOVID19}. The proposed interactive model uses location data of mobile devices, COVID-19 cases data, and population census data to compute the social distancing index of the US states or counties. The computed social distancing index provides an estimate of how likely the people of a state or a county are obeying the government order about social distancing and hence helps to reduce the spreading of the disease in that region. The location data representing the movement of human and vehicles are used in different deep learning algorithms to impute or infer other mobility-related data such as travel mode (air, car, bus, walking, etc.), trip length, trip purpose, points-of-interest visited (restaurants, shops), and socio-demographics of the travelers (income, gender, race, etc.). The imputation shows more than 90\% accuracy. The computational algorithms are also validated based on a variety of independent datasets such as the National Household Travel Survey and the American Community Survey. The resultant datasets are daily updated and publicly available. 

The Johns Hopkins Coronavirus Resource Center (CRC), through their interactive platform \cite{JohnsHopkins2020} continuously updates the sources of COVID-19 data and expert guidance. To inform the public, policy makers, and healthcare professionals about the COVID-19 pandemic and to respond accordingly, the platform aggregates and analyzes the COVID-19 cases, testing, contact tracing, and vaccine efforts. The COVID-19 project model of Institute for Health Metrics and Evaluation (IHME) \cite{IHME2020} is perhaps the most visited COVID-19 prediction model in the US. In response to requests from the University of Washington School of Medicine and other US hospital systems and state governments, IHME's COVID-19 projections were developed to determine when COVID-19 would overwhelm their ability to care for patients. IHME's COVID-19 forecast model shows the daily demand for hospital services, daily and cumulative deaths due to COVID-19, rates of infection and testing, and the impact of social distancing measures organized by county and by state. 

These interactive platforms are helping policy makers, health professionals, and researchers to understand the interplay between different factors of COVID-19 spread and undertake appropriate policy measures to control pandemic situations and improve quality of lives.

\subsection{Applications of ICT and ML techniques for COVID-19 surveillance}
Information and communication technologies (ICTs) are popularly being used in different places around the world to monitor COVID-19 situations (Table \ref{tab:mobilityDataML}). To facilitate human movement during the pandemic, some studies have explored possible options for transferring people safely from origin to destination. For example, Darsena \textit{et al.} \cite{darsena2020safe} proposed a Safe and Reliable Public Transportation Systems (SALUTARY) system to proactively tackle crowding situations in public transportation (PT) during the COVID-19 pandemic using the technologies of the Internet of Things (IoT). The system is proposed for adoption in the various segments of the public transportation system (buses/trams/trains, railway/subway stations, and bus stops) to (i) monitor and predict crowding events, (ii) adapt public transport system operations in real time (i.e., modifying service frequency, timetables, routes, and so on), and (iii) inform the users by electronic displays installed in correspondence of the bus stops/stations and/or by mobile transport applications. Another study used a ML driven intelligent approach to trace daily train travelers in different age cohorts of 16-59 years and over 60 (i.e., vulnerable age-group) to recommend certain times and routes for safe traveling \cite{Asad2020Travelers}. The study utilized the ICTs, including WiFi, RFID, Bluetooth, and UWB (Ultra-WideBand). The LUO (London Underground and Overground) Network dataset was used and various ML algorithms were compared to correctly classify different age group travelers. The results of the models indicate that the Support Vector Machine (SVM) performs better to predict the mobility of travelers with an accuracy of up to 86.43\% and 81.96\% in the 16-59 and over 60 age groups, respectively. 

Researchers in \cite {bao2020covid} estimated the changes in people mobility during the pandemic and local policy measures using a Spatio-temporal Generative Adversarial Network (COVID-GAN), a conditional generative neural network approach. They collected and integrated data from multiple sources (e.g., SafeGraph, US Census, CDC, and local government) for the city of Boston, US to provide multi-view insights in estimating mobility changes. The experimental results show that COVID-GAN sufficiently mimics real-world scenarios and performed reasonably well for a spatially unseen region (i.e., regions are not included in the training dataset) that is relatively small and is still adjacent to the spatially seen data. Similarly, this method can maintain a high quality to predict human mobility for the unseen periods, which is important to evaluate the effects of ongoing COVID-19 related policies.

Ahmed \textit{et al.} \cite {ahmed2020deep} have detected violation of social distancing measures using a deep learning-based social distancing monitoring framework using the YOLOv3 object recognition paradigm in video sequences. Moreover, this approach used the transfer-learning method to enhance model performance. The experiment results show that the developed framework efficiently identifies the people who walk too close and violate social distancing measures with an accuracy of 92\% and 98\% without and with transfer learning, respectively. The model has a tracking accuracy of 95\%. Thus, video surveillance using emerging technologies (e.g., Bluetooth, smartphones, global positioning systems, computer vision, ML, and deep learning) are very effective and provide critical solutions for enforcing social distancing measures during emergencies. Research has shown that applications of ICT and ML-based approaches are crucial to direct people on how to travel during this pandemic situation.

\section{Urban Air quality}
Besides mitigating COVID-19 transmission and associated risks, confinement and lockdown measures yield some additional health benefits by inducing a drop in air pollutants \cite {lian2020impact, kumari2020impact, singh2020impact, lenzen2020global, pickering2020identifying}. Researchers estimated that emission of ${PM_{2.5}}$, ${SO_2}$, and ${NO_2}$ was reduced by 2.5 Gt, 0.6 Mt, and 5.1 Mt, respectively in the world from the start of the pandemic through May 2020 \cite{lenzen2020global}. Many studies empirically investigated the impacts of COVID-19 related confinement and lockdown measures on air quality in urban areas specifically (Table \ref{tab:airQGeoContext}). The main results of these studies are reported in Table \ref{tab:pollutant summary}. Some of these recent studies have found that social distancing measures can improve air quality significantly \cite {zambrano2020indirect, zhu2020mediating, liu2021effects}. 
Researchers estimated $NO_2$ reduction of 22·8 $\mu g/m^3$ in Wuhan and 12·9 $\mu g/m^3$ in 367 cities of China \cite {chen2020air}. It is also reported that $PM_{2.5}$ was reduced by 1·4 $\mu g/m^3$ in Wuhan and 18·9 $\mu g/m^3$ across 367 cities of China.
NASA’s scientists reported a 30\% reduction in $NO_2$ emissions in Central China \cite {dutheil2020covid}. The same study also estimated 25\% and 6\% reduction in $CO_2$ emissions in China and worldwide, respectively. About 50\% reduction of $PM_{2.5}$ and $PM_{10}$ were observed in New Delhi, India after the implemention of lockdown measures \cite {kotnala2020emergence} in March 2020. Another study in India \cite {kumari2020impact} reported reduction in $PM_{10}$, $PM_{2.5}$, $NO_2$ and $SO_2$ concentration in Delhi (55\%, 49\%, 60\%, and 19\%, respectively) and Mumbai (44\%, 37\%, 78\% and 39\%, respectively). A 36\% and a 51\% reduction in $PM_{2.5}$ and $NO_2$ concentrations, respectively, were noted shortly after the shut down of New York City \cite {zangari2020air}. Quito, Ecuador, also reported  5.6 times and 4.8 times lower concentration of $NO_2$ in 2020 compared to 2018 and 2019, respectively \cite {zambrano2020has}. Another study from South America reported a drastic reductions in the concentrations of $NO$ (up to 77.3\%), $NO_2$ (up to 54.3\%), and $CO$ (up to 64.8\%) in the State of São Paulo, Brazil \cite {nakada2020covid}. Therefore, it is evident from recent studies that urban air quality improved substantially due to citywide lockdown measures and travel cutbacks from local populations and businesses.

The reduction in criteria pollutants in the atmosphere indirectly saved human lives. Researchers in \cite {dutheil2020covid} mentioned that improved air quality during the quarantine period avoided a total of 8,911 $NO_2$-related deaths in China (i.e., about 6\% of normal deaths in China). Another study \cite{granella2020covid} estimated that improved air quality has saved at least 19\% of premature deaths because of regular respiratory illness and 11\% year of life lost. Moreover, a number of studies also reported an increase in $O_3$ due to the reduction of other pollutants (i.e., $NO_2$, $PM_{2.5}$) \cite {zambrano2020has, nakada2020covid, broomandi2020impact}. For example, researchers in \cite {nakada2020covid} estimated a 30\% increase in ozone concentration in the urban areas of the State of São Paulo, Brazil. Despite a significant decrease in the concentration of primary pollutants (i.e., $SO_2$ 5-28\%, $NO_2$ 1-33\%, $CO$ 5-41\%, $PM_{10}$ 1.4-30\%), the concentration of $O_3$ increased by 0.5-103\% in Iran \cite {broomandi2020impact}. Thus, improving air quality has associated benefits which points to the need to undertake an integrated policy option to reduce air pollution in the urban areas to protect public health.

\begin{figure*}[htbp]
    \centering
    \includegraphics[width=\linewidth]{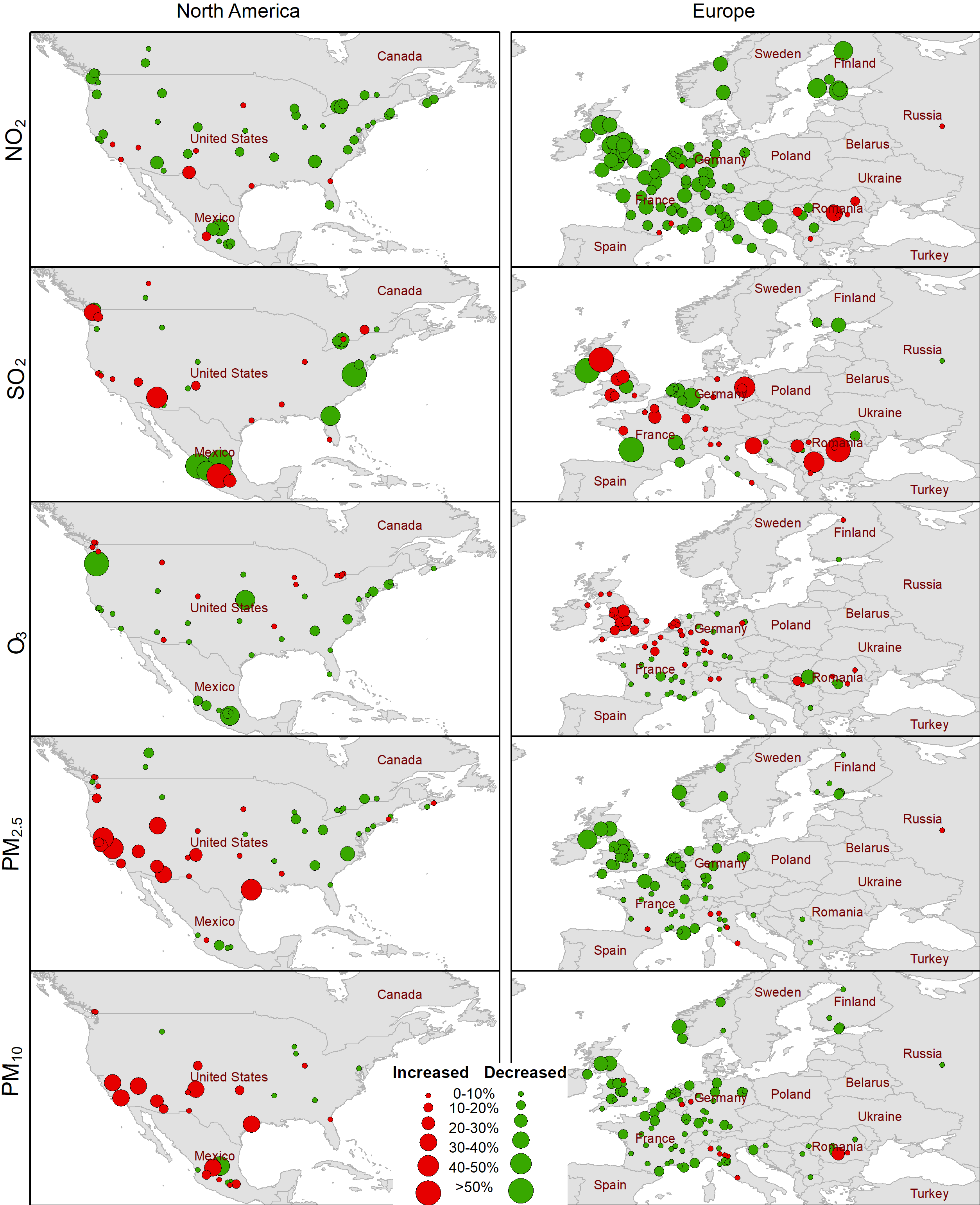}
    \caption{Daily mean air quality change in 2020 compared to 2019 in the US and Europe.
    Data source: EPA and World Air Quality Index Project, available at \cite{world_air_quality_index}.}
    \label{Fig:air_quality}
\end{figure*}

Fig. \ref{Fig:air_quality} illustrates daily mean air quality changes in 2020 compared to 2019 in North America and Europe. The changes in five criteria pollutants in major cities are mapped to understand the overall impacts of lockdown measures, travel cutbacks, and pause in industrial production on air quality. Many recent studies confirmed that air quality improved due to these factors overall. However, as Fig. \ref{Fig:air_quality} indicates, quality improvement may vary across pollutants and geographic locations. The concentration of ${NO_2}$ decreased significantly in most of the cities in Europe and North America. It is observed that the pandemic seems to have a limited and no impact on ${SO_2}$ concentration. An opposite scenario can be observed for ${O_3}$ concentration which is increased significantly in most cities. This is probably because of the fact that ${O_3}$ formation depends on the availability of other pollutants. The concentration of particulate matters mostly decreased in cities of Europe, whereas it is increased in most of the mid-eastern cities in the USA. Although, there could be many factors associated with air quality, Fig. \ref{Fig:air_quality} indicates the impact of the level of various aspects of the pandemic, including lockdown measures. Europe was under much stricter lockdown measures compared to North America. As a result, most of the criteria pollutants are shown to have decreased in Europe.

\begin{table*}[]
\centering
\caption{Geographical context and objectives of past studies on COVID-19 and air quality.}
\label{tab:airQGeoContext}
\begin{tabular}{|p{1cm}|p{2.5cm}|p{13cm}|}
\hline
\textbf{Author} & \textbf{Study context} & \textbf{Objective} \\ \hline \hline
\cite {luo2020distribution} & US & Investigating the impacts of environmental, health, socioeconomic, and demographic risk factors on COVID-19 death rates. \\ \hline
\cite   {Linka2020Reopening} & Labrador, Newfoundland, Canada & Exploring the impact of partial and total reopening of airports on COVID-19 outbreak \\ \hline
\cite   {pickering2020identifying} & US & Predicting the number of COVID-19 cases at the county level. \\ \hline
\cite   {zambrano2020has} & Quito, Ecuador & Analyzing the effect of quarantine policies on air quality (i.e., the concentration of $NO_2$, $PM_{2.5}$, and $O_3$). \\ \hline
\cite{granella2020covid} & Lombardy, Italy & Investigating the effect of lockdown measures on $PM_{2.5}$ and $O_3$ concentration. \\ \hline
\cite{mallik2020prediction} & Kanpur, India & Prediction and analyze the spatial distribution of $PM_{2.5}$ during various phases of lockdown for COVID-19. \\ \hline
\cite{lovric2020understanding} & Graz, Austria & Investigating the effect of lockdown measures on air quality ($NO_2$, $PM_{10}$, $O_3$, and Total Oxidant). \\ \hline
\cite   {keller2020global} & 46 countries of the world & Investigating the effect of lockdown measures on $NO_2$ and $O_3$ emission. \\ \hline
\cite{petetin2020meteorology} & Spain & Measuring the impact of lockdown measures on $NO_2$ emission. \\ \hline
\cite{velasquez2020gaussian} & Lima, Peru & Exploring the effect of air pollution on COVID infection. \\ \hline

\cite   {sethi2020monitoring} & Delhi, India & Monitoring the effect of lockdown on the various air pollutants during the coronavirus   pandemic. \\ \hline
\cite   {rahman2020air} & Dhaka, Bangladesh & Studying the impact of lockdown scenarios on air quality and COVID-19 transmission. \\ \hline
\cite   {tadano2020dynamic} & Sao Paolo, Brazil & Investigating the impacts of lockdown measures under four hypothetical scenarios (i.e., 10\%, 30\%, 70\%, and 90\% lockdown) on air pollution levels ($CO$, $O_3$, $NO_2$, NO, $PM_{2.5}$, and $PM_{10}$) \\ \hline
\cite   {magazzino2020relationship} & Paris, Lyon, and Marseille, French & Investigating the relationship between air pollution and  COVID-19   transmission. \\ \hline
\cite {mirri2020covid} & Emilia-Romagna, Italy & Predicting the possibility of resurgence (second wave) of the COVID-19 pandemic \\ \hline
\cite   {xu2020variation} & Hangzhou, China & Estimated the impacts of COVID-19 lockdown measures on Black carbon (BC) emissions \\ \hline
\cite {cole2020impact} & 30 cities in China & Quantifying the impact of COVID-19 lockdown on the concentration of air pollutants. \\ \hline
\cite {gatti2020machine} & Italy & Evaluating the effects of exposure to air pollutants on COVID-19 susceptibility. \\ \hline
\cite   {guevara2020time} & 30 European countries & Quantifying the reduction of primary pollutants from the energy industry, manufacturing industry, surface, and air transportation during COVID-19 lockdowns. \\ \hline

\cite {mele2020pollution} & 25 major cities, India & Exploring the relationships between pollutants emission, economic growth, and COVID-19 related deaths \\ \hline
\cite   {vito2020high} & Portici City, Italy & Monitoring air quality during COVID-19 pandemic with the help of IoT intelligent multisensory devices. \\ \hline
\cite{wang2020four} & 6 Megacities, China & Analyzing the impacts of COVID-19 related lockdown measures on air quality. \\ \hline
\cite   {barre2020estimating} & 100 European cities & Estimating changes in NOx due to COVID-19 lockdown. \\ \hline
\end{tabular}
\end{table*}

\begin{table*}[]
\caption{Changes in air pollutants induced by COVID-19 related lockdown measures ('-' sign indicates a decrease and '+' sign indicates an increase in pollutants).}
\label{tab:pollutant summary}
\centering
\begin{tabular}{|p{1cm}|p{2.8cm}|p{3.3cm}|p{4.7cm}|p{1.5cm}|p{1.9cm}|}
\hline
\textbf{Study} & $\mathbf{CO}$, $\mathbf{CO_2}$ & $\mathbf{NO}$, $\mathbf{NO_2}$ & $\mathbf{PM_{2.5}}$, $\mathbf{PM_{10}}$ & $\mathbf{SO_2}$ & $\mathbf{O_3}$ \\ 
\hline \hline
\cite{granella2020covid} & - & -33\% $NO_2$ & -15\% $PM_{2.5}$ & - & -  \\ \hline
\cite{mallik2020prediction} & - & - & -4\% (for lockdown), -47\%  for post lockdown with previous year & - & -  \\ \hline
\cite{lovric2020understanding} & - & -(36.9 - 41.6)\% & -(6.6 - 14.2)\%& - & + (11.6 - 33.8)\% \\ \hline

\cite{keller2020global} & - & -18\% & - & - & +50\%  \\ \hline
\cite{petetin2020meteorology} & - & -50\% $NO_2$ & - & - & - \\ \hline

\cite{velasquez2020gaussian} & - & -36\% $NO_2$ & Insignificant change & - & -  \\ \hline
\cite{sethi2020monitoring} & $CO$ Decreased & $NO_2$ Decreased & Both Decreased & Decreased & Increased  \\ \hline

\cite{rahman2020air} & -8.8\% $CO$ & -20.4\% $NO_2$ & -26\% $PM_{2.5}$ & -17.5\% & -9.7\%   \\ \hline
\cite{tadano2020dynamic} & Decreased $CO$ & Decreased NO \& $NO_2$ & Decreased $PM_{2.5}$ \& $PM_{10}$ & - & Decreased  \\ \hline

\cite{cole2020impact}& Insignificant $CO$ change & -63\% $NO_2$ (fell by 24ug/m3) & -63\% $PM_{10}$  & Insignificant change & -  \\ \hline
\cite{guevara2020time} & - & -33\% $NO_2$ & -7\% $PM_{2.5}$ & -7\% & -  \\ \hline

\cite{vito2020high}& -(7-11)\% $CO$ & - (23 - 37)\% $NO_2$ & -(14 - 20)\% $PM_{10}$, -(7 - 16)\% $PM_{2.5}$ & -(2 - 20)\% & -  \\ \hline
\cite{wang2020four} & $CO$ decreased  & -(36-53)\% $NO_2$ & $PM_{2.5}$ Decreased  & - & Decreased  \\ \hline

\end{tabular}
\end{table*}

\subsection{Impact on concentration of Particulate Matters (PM)}
Particulate Matters (PM), composed of solids and gases, are largely generated during the burning of fossil fuels and woods in the transportation, manufacturing industry, and power plants \cite {epa2020covid19}. Due to lockdown measures, human mobility is reduced, and the majority of industries are closed; hence, emissions of PM have been reduced substantially. Some researchers have investigated the level of PM in the air during lockdown periods and compared the results to pre-COVID-19 situations to assess the changes in air quality. For example, Mallik \textit{et al.} \cite{mallik2020prediction} estimated the concentration of \(PM_{2.5}\) in Kanpur city, India for three different lockdown phases: pre, during, and post lockdown conditions in relation to the pandemic. They used four different ML approaches to estimate \(PM_{2.5}\) from remote sensing-based MODIS Aerosol Optical Depth (AOD) data and meteorological parameters, including temperature, rainfall, relative humidity, wind speed, and mixing height. A hybrid ML approach that combines ANN and Multiple Linear Regression (MLR) shows the highest performance for the estimation of \(PM_{2.5}\). The hybrid approach outperformed (\(R^2=0.96\)) Linear Regression (LR) (\(R^2=0.016\)), MLR (\(R^2=0.246\)) and ANN (\(R^2=0.895\)) models. The authors compared the estimated \(PM_{2.5}\) of this year with the previous year and reported 4\% and 47\% reduction in \(PM_{2.5}\) during the lockdown and post-lockdown condition, respectively. From the analysis, they concluded that \(PM_{2.5}\) is reduced because of the reduction in emissions from industries and transport vehicles.

\begin{figure}[ht]
    \centering
    \includegraphics[width=.95\linewidth]{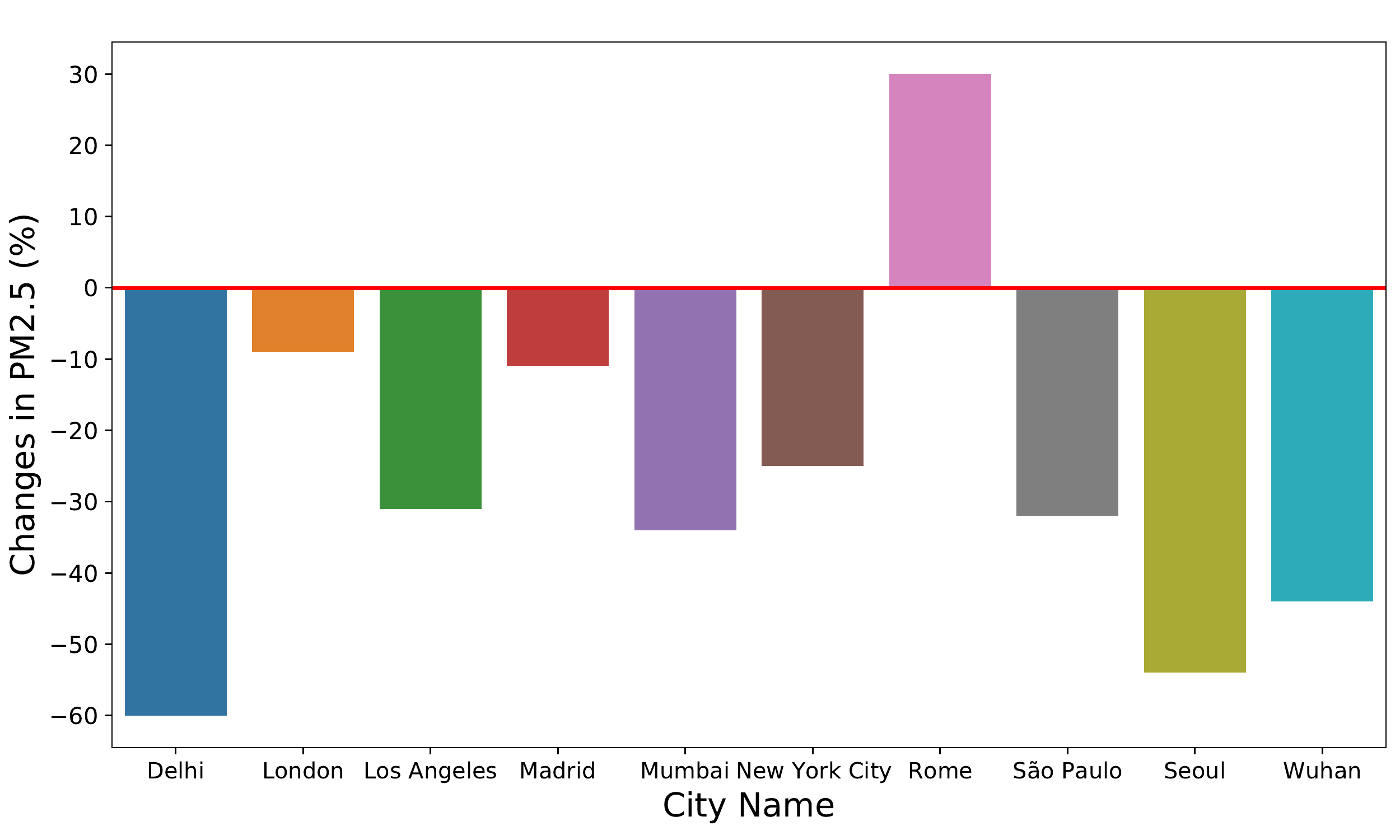}
    \caption{Changes in $PM_{2.5}$ levels in 2020 compared to 2019 in 10 major global cities during COVID-19 related lockdown periods between February and April 2020 \cite {air2020world}.}
    \label{Fig:PM2.5}
\end{figure}

The IQAir \cite {air2020world} analyzed $PM_{2.5}$ levels in 10 major and severely affected cities in the world by observing three weeks of the strictest lockdown conditions in-between February and April in 2020 (Fig. \ref{Fig:PM2.5}). The figure indicates that 9 out of 10 global cities have experienced a reduction in $PM_{2.5}$ during the mentioned period in 2020 compared to 2019. The report also mentioned that the cities with higher levels of $PM_{2.5}$ concentration achieved the most substantial reductions (e.g., Delhi, Seoul, Wuhan) due to restriction on vehicular movement, closure of the educational institutions, industries, and workplaces, and shutdown of all non-essential businesses. In contrast, Rome showed a 30\% increase in $PM_{2.5}$ concentration due to a higher reliance on residential heating systems, and the shift away from public transportation to private cars.

Simulating four hypothetical scenarios (i.e., 10\%, 30\%, 70\%, and 90\% of lockdown), Tadano \textit{et al.} \cite{tadano2020dynamic} predicted air pollution levels ($PM_{2.5}, \ and \ PM_{10}$) in São Paulo, Brazil. Various data, including the daily number of COVID-19 cases, partial lockdown level, and meteorological variables, were feed into four ANN models (Multilayer Perceptron, Radial basis function, Extreme Learning Machines, Echo State Networks) to simulate pollutant levels. The result shows that Multilayer Perceptron (MLP) outperformed other ANN models. The model results reported a decrease in the concentration of $PM_{2.5}, \ and \ PM_{10}$. It is also evident that there is a negative correlation between lockdown levels and air pollutants (i.e., a higher level of lockdown measures reduces the amount of PM in the air to a greater extent). 

Grabekka \textit{et al.} \cite{granella2020covid} has investigated the effect of lockdown on \(NO_2\) because of the dramatic decrease in mobility and economic activities in Lombardy, Italy. For that purpose, they built a synthetic counterfactual estimation of air quality based on meteorological variables using the Extreme Gradient Boosting Regressor. The model was trained by the atmospheric data between 2012 and 2019 from 227 weather stations, where, meteorological variables are the predictor variables and air qualities are the predicted variables. Like other studies, they also considered calendar variables such as the day of the year to capture the trend over time. Unlike other studies, they utilized a ratio between \(PM_{2.5}\) and \(PM_{10}\) as an additional predictor variable to avoid the impact of the pollutants transported from a long distance, such as mass dust from the Caspian Sea. The correlation value over 0.87 between observed and predicted values before lockdown indicates the strength of the model. The result shows that the lockdown reduced the concentration of \(PM_{2.5}\) and \(NO_2\) by 3.84 $\mu g/m^3$ (16\%) and 10.85 $\mu g/m^3$ (33\%), respectively. 

Some other studies also reported the role of lockdown measures for the reduction of $PM_{2.5}$ concentration in the atmosphere \cite{vito2020high, wang2020four}.  A study estimated a reduction of $PM_{2.5}$ by 7\% in European countries \cite{guevara2020time}. Similarly, another study shows an average of 14.2\% reduction in \(PM_{10}\) compare to estimated concentration in 2020 and 11\% reduction compared to the historical mean value between 2014 and 2019 \cite{lovric2020understanding}. In a developing country context, a study reported a 26\% to 54.2\% reduction in the level of $PM_{2.5}$ concentration during full and partial lockdown scenario in Dhaka, Bangladesh \cite{rahman2020air}.

\subsection{Impacts on the concentration of nitrogen oxides $(NO_x)$}
The main source of nitrogen oxides (i.e., $NO$, $NO_2$) is burning of fossil fuel in the transportation, industries, and power plants \cite {epa2020covid19}. These highly reactive gases have detrimental effects on public health by affecting the respiratory systems of people, besides broader environmental consequences. Many studies in different geographical contexts have reported a reduction in the concentration of nitrogen oxides during the COVID-19 lockdown periods due to the limited mobility of people and the closure of workplaces and industries. Change in traffic emission during lockdown played a major role in the substantial reduction in ${NO_2}$ in six megacities of China \cite{wang2020four}. Random forest learning-based models reported a reduction of ambient ${NO_2}$ concentration of about 36-53\% during a four-month (January to April 2020) lockdown due to the formation of secondary aerosols, which enhanced the oxidizing capacity of the atmosphere. Subsequent lifting of level-1 control action caused  ${NO_2}$ to drop below 10\% in late April. 

Petetin \textit{et al.} \cite{petetin2020meteorology} estimated the business-as-usual \(NO_2\) mixing ratio in more than 50 provinces and islands across Spain and compared it with observed hourly \(NO_2\) concentration to evaluate the impact of lockdown measures in March 2020. A Gradient Boosting ML algorithm was trained on the \(NO_2\) concentration from January 1 to 23rd April 2020 and meteorological data including temperature, pressure, wind speed, cloud cover, solar radiation, ultra violet radiation, and calendar variables. The model shows reliable performance based on the model uncertainty assessment where overall bias, root mean square error, and correlation are +4\%, 29\%, and 0.86, respectively. The result shows an overall 50\% reduction in $NO_2$ due to a reduction in road and air transportation during lockdown periods. Using a similar ML method, a study \cite {barre2020estimating} estimated the reduction in $NO_2$ emission to be 23\% to 43\% in 100 European cities. They also mentioned that the cities with strict lockdown measures have experienced a stronger reduction in $NO_2$. Another study in a similar context (i.e., city of Portici, Italy) \cite{vito2020high} used multilinear regression (MLR) model and a shallow neural network (SNN) model to monitor air quality with the help of IoT intelligent multisensor devices during phase 2 of the pandemic. Results suggested that due to the reduction of mobility, particularly from car travel during the lockdown period, the concentration of ${NO_2}$ has dropped significantly. 

The study in \cite{guevara2020time} has used gradient boosting machine (GBM) models to quantify the diminution of primary pollutants from the energy industry (e.g., power plants, heat plants), manufacturing industry, surface and air transportation in 30 European countries during the COVID-19 lockdowns. Data collected from different sources dated from 21 February to 26 April 2020 and historical data of the previous 5 years (2015-2019) during the same time period to assess the level of pollutants in the air. The results indicate that severe lockdown at the EU-30 level countries caused average emission reduction of ${NO_x}$, Non-methane volatile organic compounds (NMVOC), and ${SO_x}$ by 33\%, 8\%, and 7\%, respectively. However, the steepest reduction in ${NO_x}$ (50\%) can be imputed to the closure of the transportation sector, which is responsible for about 85\% reduction of all pollutants overall. It is also evident that the drop of ${NO_2}$ reached up to 58\% in the urban areas, whereas it was only 44\% in rural areas; the average contribution of the transportation sector was 86\% and 96\%, respectively.

Another approach to investigate the impact of lockdown measures on air quality is to estimate air pollutants for 2020 based on the trend of previous years and to compare it with the true values of these pollutants. This approach generates a synthetic estimate of what air quality would have been without a lockdown effect based on historical data. Lovric \textit{et al.} \cite{lovric2020understanding} investigated the lockdown effect on air quality at five monitoring stations in Graz, Austria. Four different pollutants ($NO_2, O_3, PM_{10}, \ and \ O_x$) were estimated based on meteorological data series, including air temperature, precipitation, wind speed, wind direction, and air pressure. The Random Forest algorithm was trained by the historical meteorological variables and pollutants data between 2014 and 2019. Based on the relationship between meteorological variables and pollutants level, air quality was estimated for the lockdown period. Finally, the estimated values were compared with observed values at the five monitoring stations to evaluate the impact of lockdown on air quality. The results show an average reduction of 36.9\% in \(NO_2\) due to the lockdown effect. Authors also compared the air quality of 2020 with the historical mean between 2014 and 2019, which also indicates a 38.1\% reduction in \(NO_2\).

Keller \textit{et al.} \cite{keller2020global} developed a bias-corrected model (BCM) for the NASA global atmospheric composition model (GEOS-CF) using the XGBoost ML algorithm. They utilized this BCM model to estimate \(NO_2\) from eight meteorological parameters, including surface wind components, surface temperature, relative humidity, cloud coverage, precipitation, pressure, and planetary boundary layer. The ML predictor was trained on 2018-2019 data to predict the model bias for the observation sites in 2020 to adjust the predicted concentrations from the GEOS-CF model. This study estimated \(NO_2\) at 5756 observation sites in 46 countries across the globe between January to June 2020. They found that on average \(NO_2\) concentrations were 18\% lower than the business-as-usual level. 

Similarly, Grabekka \textit{et al.} \cite{granella2020covid} found that the lockdown reduced the concentration of \(NO_2\) by 10.85 $\mu g/m^3$ (33\%). A 20.4\% to 55.5\% reduction in $NO_2$ during full and partial lockdown scenario, respectively, compared to the pre-COVID situation, was also reported in Dhaka, Bangladesh \cite{rahman2020air}. Thus, it is evidenced that the concentration of $NO_x$ has reduced significantly in urban areas due to city-wide and countrywide lockdown measures and closure of transportation and workplaces \cite{tadano2020dynamic}.

\subsection{Impacts on the concentration of $CO_x$}
Similar to nitrogen oxides, carbon oxides (i.e., $CO$, $CO_2$) are also released by burning fossil fuel by  vehicles and industries, and by biomass \cite {epa2020covid19, xu2020variation}. These colorless gases can negatively affect human health when present at a high level of concentration in the air. However, studies have found that the level of carbon emission dropped significantly during the COVID-19 pandemic \cite{tadano2020dynamic, wang2020preventing}. These studies empirically investigated changes in carbon emission during the lockdown periods. For example, A study in Dhaka, Bangladesh, \cite{rahman2020air} investigated the impact of different lockdown scenarios on air quality and on COVID-19 transmission using data from weather monitoring stations. Using generalized additive models (GAMs), wavelet coherence, and random forest (RF) models, this study found a 8.8\% to 23.5\% reduction in CO concentrations during full and partial lockdown scenarios, respectively, compared to the pre-COVID situation. Similarly, this study reported a 17.5\% to 48.1\% reduction in $SO_2$ during full and partial lockdown scenarios, respectively, compared to the pre-COVID situation. 

\begin{figure}
  \begin{subfigure}[b]{0.5\textwidth}
    \includegraphics[width=.95\linewidth]{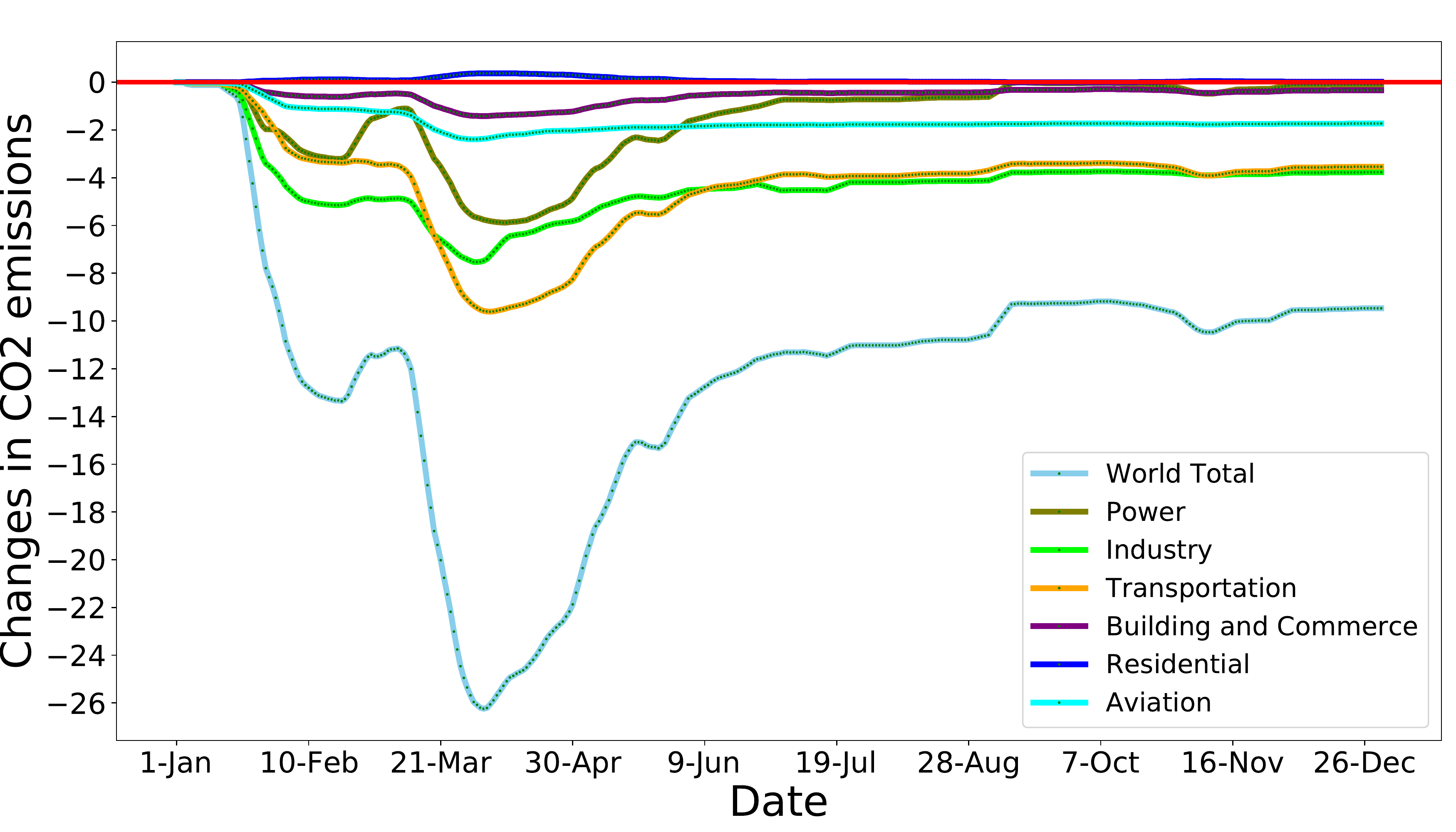}
    \caption{$CO_2$ emissions under high level of confinement.}
    \label{Fig:CO2_high}
  \end{subfigure}
  \begin{subfigure}[b]{0.5\textwidth}
    \includegraphics[width=.95\linewidth]{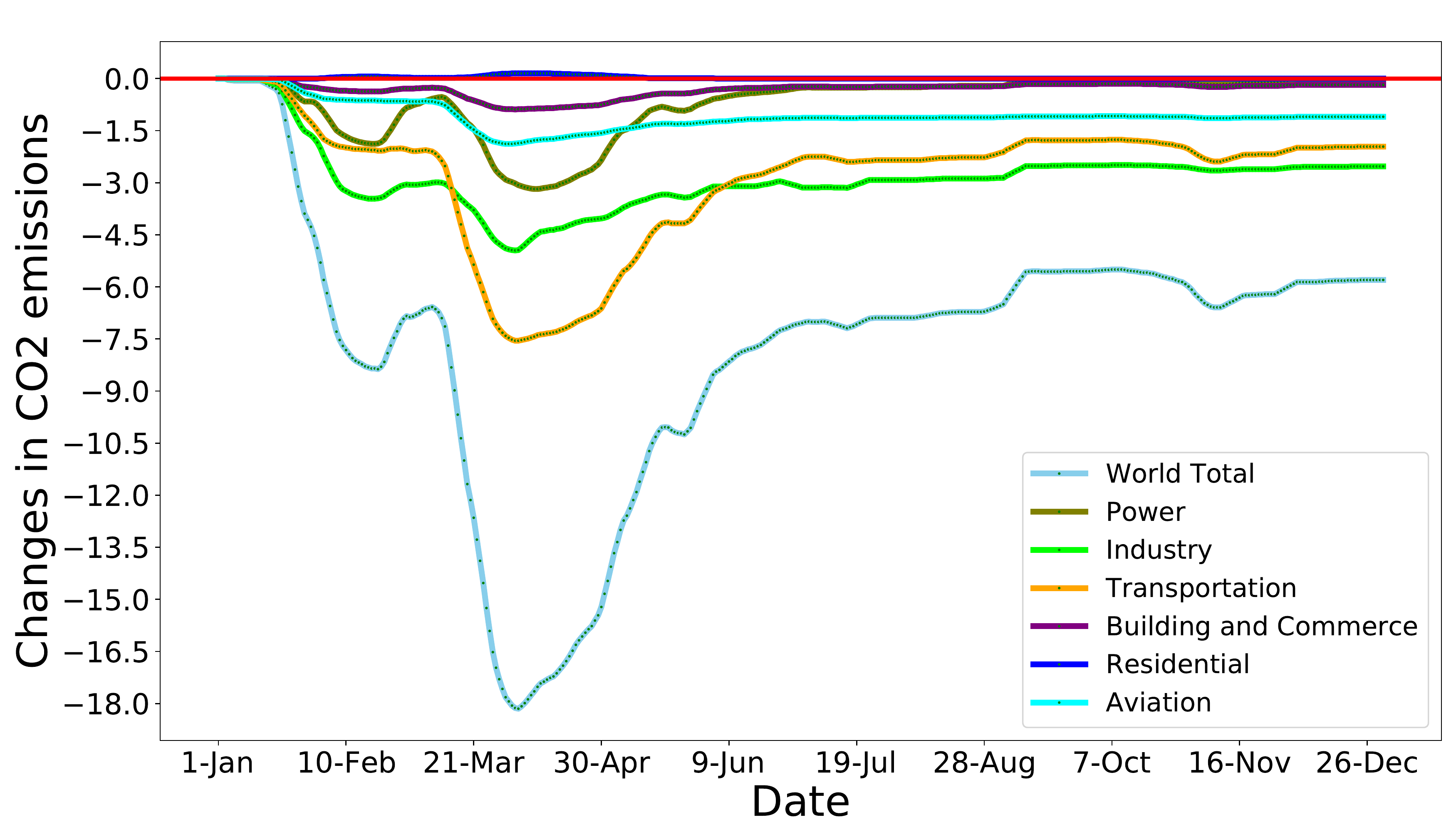}
    \caption{$CO_2$ emissions under medium level of confinement.}
    \label{Fig:CO2_med}
  \end{subfigure}
    \begin{subfigure}[b]{0.5\textwidth}
    \includegraphics[width=.95\linewidth]{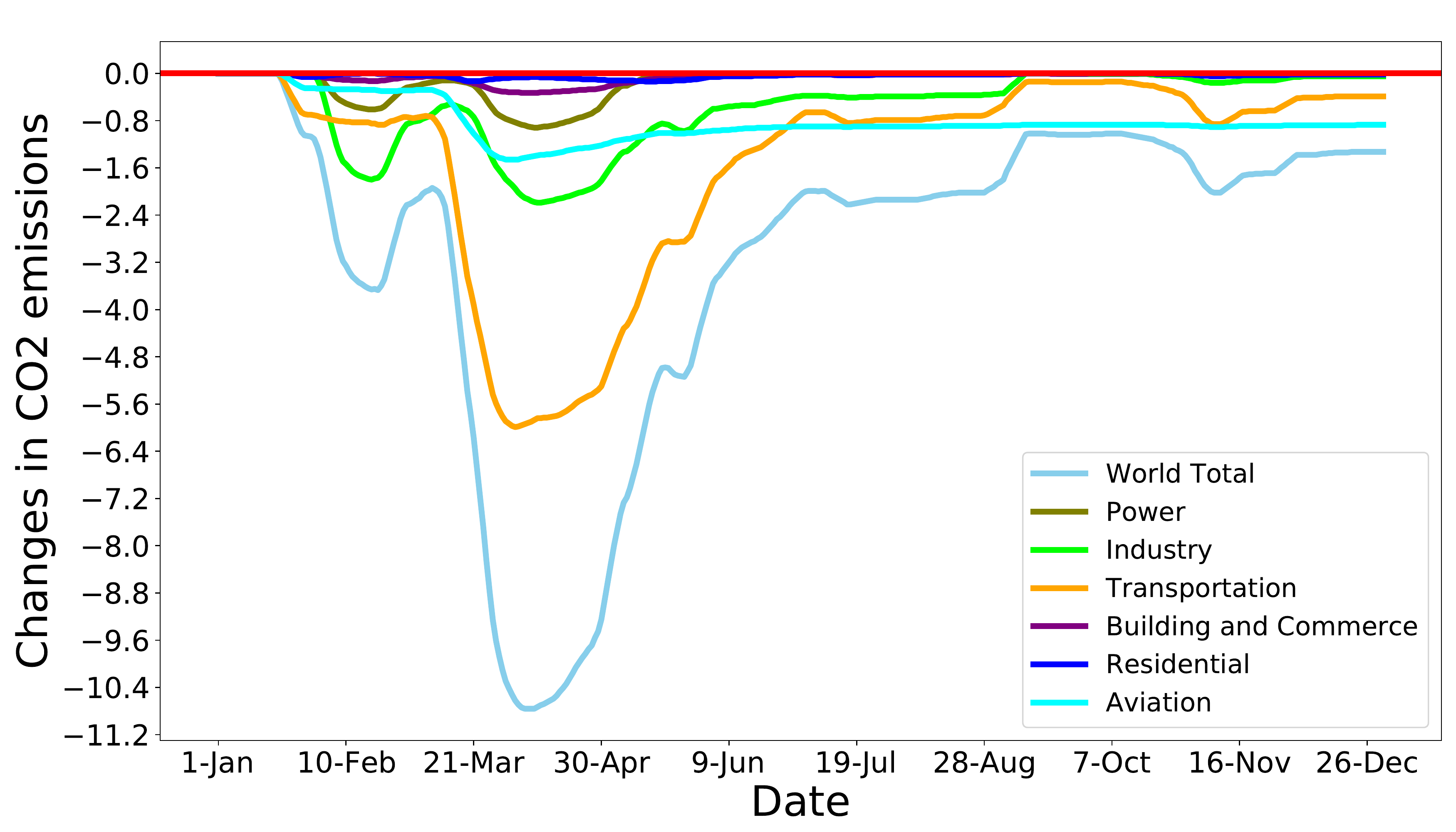}
    \caption{$CO_2$ emissions under low level of confinement.}
    \label{Fig:$CO_2$_low}
  \end{subfigure}
  %
   \caption{Longitudinal changes in $CO_2$ emissions in the world under confinement scenarios.}
    \label{Fig:CO2_emission}
\end{figure}

After collecting data on energy, human activities, and policy measures, researchers in \cite {le2020temporary} estimated daily changes in carbon emissions in 2020 compared to 2019 for three-level of confinement (low, medium, and high) and for six sectors of the economy (e.g., power, industry, surface transportation, public building and commerce, residential, and aviation). Fig. \ref{Fig:CO2_emission} shows the total changes in ${CO_2}$ emission in the world as a whole, as well as the changes in the six economic sectors under three confinement scenarios for the period of 01 January 2020 to 31 December 2020. The figure demonstrates a positive association between COVID-19 related confinement and changes in $CO_2$ emission (i.e., a higher level of confinement reduces further $CO_2$ emissions from all economic sectors). The peak reduction in $CO_2$ emissions is evidenced in March and April due to the severity of the pandemic, substantial mobility reduction, and associated fear and anxiety. However, the reduction in $CO_2$ emissions from transportation and industry is relatively greater than other sectors due to strict restrictions on mobility, closure of workplaces and industries.

Researchers in \cite {xu2020variation} estimated the impacts of COVID-19 lockdown measures on black carbon (BC) emissions in urban and urban‐industry areas, suburbs, and rural areas of Hangzhou, China using a multiwavelength Aethalometer model. Analyzing data collected from nine observation sites, the study showed a city-wide 44\% reduction in BC from 2.30 to 1.29 $\mu g/m^3$ due to reduction of vehicle emissions in the urban areas and biomass burning in the rural areas during the COVID‐19 lockdown periods. More precisely, this study reported 47\%, 49\%, 41\%, and 38\% BC emission in urban, urban industrial, suburban, and rural areas, respectively. Using a similar method, researchers in \cite {goel2020variations} found reductions of 78\%, 67\%, 53\%, 59\%, 74\%, and 66\% in BC during Lockdown-1 (25th March-14th April, 2020), Lockdown-2 (15th April-3rd May, 2020), Lockdown-3 (4th-17th May, 2020), Lockdown-4 (18th-31st May, 2020), Unlock-1 (June 2020), and Unlock-2 (July 2020) situations, respectively compared to the pre-Lockdown period (18th Feb-24th March, 2020). Using Concentration Weighted Trajectories (CWT) analysis, this study concluded that local sources (e.g., fossil fuel, biomass burning) were primarily responsible for BC concentration over Delhi, India. 

Thus, the closure of public transport, workplace, and industry due to government-mandated lockdown measures significantly reduces carbon emission by decreasing the burning of fossil fuel, solid substances, and biomass in urban and rural areas \cite {chen2020impact, peng2020brown}.

\subsection{Impacts on the concentration of $O_3$}
Unlike the other pollutants, most of the studies have mentioned that the concentration of $O_3$ has increased dramatically during the COVID-19 lockdown periods \cite{tadano2020dynamic, shi2020response, sicard2020amplified, siciliano2020increased}. Experts mentioned that the reduction in traffic emission is largely responsible for increasing ${O_3}$ concentrations \cite{wang2020four}. For instance, researchers in \cite{lovric2020understanding} observed that $O_3$ concentration was increased significantly between 11.6\% to 33.8\% during the lockdown period in Graz, Austria, which can be explained by the reduction of the \(NO-O_3\) titration cycle when \(NO_x\) emission was low during the lockdown. Similarly, \(O_3\) concentration was increased by up to 50\% in some locations of the world where \(NO_2\) was also reduced significantly \cite{keller2020global}. However, the response of \(O_3\) is nonlinear to the reduction of $NO_2$. Some cities such as Beijing and Madrid show very little difference in \(O_3\), while having declines in \(NO_2\) comparable with other cities. The study in Dhaka, Bangladesh \cite{rahman2020air} also observed a 9.7\% to 31\% reduction in $O_3$ concentration during full and partial lockdown scenarios, respectively, compared to the pre-COVID situation. Thus, it can be concluded that the overall level of $O_3$ increased significantly during the lockdown periods in many parts of the world.

\subsection{Impacts of air pollutants on COVID-19 transmission and death rate}
According to public health researchers, indoor and outdoor air quality significantly affects human health \cite {wolkoff2018indoor, schweitzer2010neighborhood, jones1999indoor}. Evidence indicates that prolonged exposure to air pollution could increase the risk of viral infections, and respiratory illness \cite{gatti2020machine}. Moreover, small particles in the air may facilitate the viral spread like coronavirus. Thus, the scientific community has grown more interested in knowing the effect of air quality on COVID-19 transmission. At the beginning of the COVID-19 pandemic, many scientists investigated the role of weather variables (e.g., temperature, pressure, precipitation, humidity) and air quality on the virus transmission rate. However, the scientist community now believes that weather conditions may have very little or no effect on virus transmission. Many studies have already investigated the relationship between local air quality and COVID-19 infection and death rate. 

A study \cite{magazzino2020relationship} explored the relationship between particulate matter ($PM_{2.5} \ and \ PM_{10}$) and COVID-19 death using ANN for three cities in France. They found different thresholds of the concentration of particulate matter responsible for fatality in different cities, for instance 17.4 $\mu g/m^3$ ($PM_{2.5}$) and 29.6 $\mu g/m^3$ $PM_{10}$ for Paris. The results indicate that an increase in $PM_{10}$ concentration beyond 29.6 $\mu g/m^3$ could generate a 63.2\% increase in mortality. Similarly, any value above 20.6 $\mu g/m^3$ in $PM_{10}$ would generate an increase in deaths of 56.12\% in Lyon, France.  
Similarly, a study in the Lima metropolitan area, Peru investigated the relationship between the COVID-19 infection rate and major air pollutants ($CO$, $NO_2$, $SO_2$, $PM_{2.5}$, and $PM_{10}$) \cite{velasquez2020gaussian}. The study used Reduced-spaced Gaussian Process Regression and ANN models. The result indicates that the industrial area shows a higher number of infections compared to other land use zones. The industrial zone features the highest air pollution because of the high concentration of $NO_2$, $PM_{2.5}$, and $PM_{10}$. 

Using various ML techniques, including Decision Trees, Linear Regression, and Random Forest, researchers in \cite{sethi2020monitoring} computed the correlation between air pollutants and COVID-19 fatality in Delhi, India. The result shows that COVID-19 fatality is positively correlated with a few air pollutants such as \(O_3\), \(SO_2\), and \(NH_3\). Moreover, results from three different ML algorithms indicate that \(NH_3\), \(NO_2\), and \(PM_{10}\) are the important factors of COVID-19 related fatalities. Similar to other studies around the world, this study also shows that there is a significant increase in ozone and toluene concentration in the air because of the lockdown measures. Thus, an increase in surface ozone may augment the fatality rates. A study in \cite{gatti2020machine} based on the Random Forest algorithm investigated the impact of air pollution, rather than direct person-to-person contamination, on the spread of SARS-CoV-2 in Italy. Interestingly, compared to lifestyle or socio-economic factors, air quality played the most significant role in pandemic diffusion and severity. A 5-10\% increase in air pollution in Italy may cause an additional rise of 21-32\% in the COVID-19 toll, having 4-14\% more deaths. These findings were achieved by analyzing epidemiological data on COVID-19 positivity, mortality, and air quality index of 20 Italian regions and 99 Italian provinces from 2015 to 2019. Furthermore, about 70\% of COVID-19 deaths nationwide might be due to emissions from industries, farms, and road traffic.

There is a unidirectional causal relationship between pollution and economic growth. In 25 major Indian cities, Mele and Magazzino \cite{mele2020pollution} investigated the relationships between economic growth, pollution emissions, and COVID-19 related deaths using time series analysis and ML based complex causality algorithm (D2C). Time-series econometric analysis indicated that there is a strong unidirectional correlation between economic growth and concentration of ${PM_{2.5}}$, ${CO_2}$, and ${NO_2}$. The ML algorithm confirmed a causal relationship between the concentration of ${PM_{2.5}}$, ${CO_2}$, ${NO_2}$, and COVID-19 deaths. Thus, air pollutants have a direct relationship with the COVID-19 pandemic-related death, with the largest influence from $PM_{2.5}$. Prolonged exposure to $PM_{2.5}$ causes acute health problems (e.g., asthma, cough, loss of cardiac or lung capacity), which results in fatalities. The results are consistent with other studies conducted in the US \cite {wu2020exposure}, where the researchers found that a 1 $\mu g/m^3$ increase in $PM_{2.5}$ concentration causes an increase in the mortality rate by 15\%. Thus, exposure to $PM_{2.5}$ aggravates the symptoms of COVID-19 with heightened risks of mortality. 


The above discussion confirms that air pollution has a significant impact on the diffusion of the COVID-19 pandemic. However, the reduction of air pollutants during COVID-19 related lockdown has had a mitigating effect, by substantially increasing air quality, which in turn had the effect of curbing the severity of the pandemic.

\subsection{AI and ML tools in air pollutants monitoring}
Nowadays, ML techniques are used extensively for their computational benefits and intelligence in ‘big-data' analytics of air pollutants monitoring. Supervised ML algorithms predict or classify data based on existing labeled data (Table \ref{tab:airQDataML}). ML algorithms are routinely used to estimate the business-as-usual concentration of criteria pollutants using large historical datasets. A number of studies also use ML algorithms to discover the relationship between air quality and COVID-19 related death, including all its nuances. Their distinctive advantage rests with the capability to handle large datasets and with their predictive reliability.

Researchers in \cite{vito2020high} used Raspberry Pi 3b+ in the IoT stationary architecture to build an air quality monitoring system. Their IoT-based model could be exploited to get a deeper understanding of the influence of air pollution on the pandemic. It can also be helpful to anticipate the increased susceptibility of individuals due to exposure and eventually to predict new outbreaks. Recently, Cole \textit{et al.} \cite{cole2020impact} have developed a meteorological normalization technique based on the random forest algorithm in combination with the augmented synthetic control method to quantify the impact of COVID-19 lockdown on air pollution and public health in Wuhan, China. Investigating the hourly data of four pollutants (${SO_2}, {NO_2}, CO, PM_{10}$) in 30 cities of China between January 2013 and February 2020, the model reported a 63\% reduction (a drop of 24 $\mu g/m^3$) of ${NO_2}$ with respect to pre-lockdown levels. Even though the study found no significant impact of lockdown on ${SO_2}$ or CO, it indicates the reduction of ${NO_2}$ could have prevented a total of 3,368 deaths in the City of Wuhan and 10,822 deaths in the whole of P.R. China.

Mirri \textit{et al.} \cite{mirri2020covid} used eight different ML models to predict the possibility of a resurgence (second wave) of the COVID-19 pandemic in all nine provinces of Emilia-Romagna, Italy for the period of September–December 2020; Emilia-Romagna was one of the most severely afflicted regions during the first phase of the pandemic from February to April 2020. They trained the models with data on COVID-19 confirmed cases from February to July 2020, and the daily measurements of $PM_{2.5}$, $PM_{10}$ and $NO_2$ in the periods of September–December 2017/2018/2019. The results from the models demonstrated that the gradient boosting model performs better, with an accuracy of 90\%, to predict the possibility of a second wave of the pandemic. Conducting a sensitivity analysis, the study commented that the use of personal protective measures has a significant impact to downgrade the likelihood of a second wave in the nine provinces of Emilia-Romagna.

In the light of the potential benefits of analysis on complex data structures with high computational efficiency, many studies have used different ML techniques to monitor air quality and explore the effects of air pollutants on disease transmission. One important question arising here is that of the motivation for using ML to predict air quality rather than statistical or econometric models. The answer may lie in the fact that these studies are investigating the effect of lockdown measures on air quality in the current situation instead of the simple comparison between the current value and historical values. Since air quality is significantly influenced by meteorological factors, a simple statistical or econometric model may not capture the true effect of lockdown on the concentration of air pollutants. Thus, machine learning to estimate air quality considering all these factors may provide a better understanding of the lockdown and confinement effects \cite{lovric2020understanding}.

\begin{table*}[]
\centering
\caption{Data and ML methods used in past studies on COVID-19 and air quality.}
\label{tab:airQDataML}
\begin{tabular}{|p{1cm}|p{11.5cm}|p{3.5cm}|}
\hline
\textbf{Author} & \textbf{Data and sources} & \textbf{ML methods} \\ \hline \hline

\cite   {luo2020distribution} & COVID-19 death cases from US Facts, Pollutants data (e.g., $PM_{2.5}$, benzene, formaldehyde, acetaldehyde, carbon tetrachloride) from Environmental Protection Agency and Centers for Disease Control and Prevention, weather (e.g., temperature, precipitation, sunlight and UV exposure), land cover, health status (e.g., disabled, obese, overweight) from Centers for Disease Control and Prevention, socioeconomics (e.g., health insurance, poverty, income) and commuting information (e.g., travel modes, time) from the US Census & Geographical weighted RF (GW-RF) \\ \hline
\cite   {Linka2020Reopening} & COVID-19 epidemiology data from \cite {berry2020open} and New York Times COVID-19, Daily air traffic (people/day) from International Air Transport Association. &  Susceptible-exposed-infectious-recovered (SEIR) models, Bayesian Interference models \\ \hline
\cite   {pickering2020identifying} & Flight data from Bureau of Transportation Statistics, ground traffic data from NYC Open Data, air pollutant data (e.g., CO, $NO_2$, Ozone, and $SO_2$) from Aura Satellite (OMI instrument) and Environmental Protection Agency & Support vector machine (SVM) \\ \hline
\cite   {zambrano2020has} & $NO_2$, $PM_{2.5}$, and $O_3$ concentrations from the Secretary of the Environment of the Municipality of the Metropolitan District of Quito & Parametric   analysis \\ \hline
\cite{mallik2020prediction} & Ground-based $PM_{2.5}$ concentration from central pollution control board (CPCB), satellite-derived MODIS Aerosol Optical Depth (AODs) data, and meteorological data (wind speed, temperature, rainfall, relative humidity, and mixing height) from Indian Meteorological Department. & Artificial neural network (ANN) \\ \hline
\cite{lovric2020understanding} & Meteorological data (e.g., temperature, precipitation, wind speed, wind direction, and air pressure), air quality  data for the year 2014-2020, and lockdown data from the Austrian government & Principal component analysis (PCA), random forest (RF) \\ \hline
\cite   {keller2020global} & Eight meteorological parameters (e.g., surface wind components, surface temperature, relative humidity, cloud coverage, precipitation, pressure, and   planetary boundary layer) from government agencies & Extreme Gradient Boosting Decision Tree (XGBDT) \\ \hline
\cite{granella2020covid} & Daily minimum and maximum temperature, average wind speed and direction, average relative humidity, daily cumulative precipitation, and $PM_{2.5}$ and $PM_{10}$   concentration from ARPA Lombardia, and time and seasonal variables & XGBDT \\ \hline
\cite{petetin2020meteorology} & Meteorological data (e.g., temperature, pressure, wind speed, cloud cover, solar radiation, ultra-violate radiation) from ERA5 reanalysis dataset, and $NO_2$ data from Earth Sciences Department of the Barcelona Supercomputing Center. & Gradient Boosting Decision Tree (GBDT) \\ \hline
\cite{velasquez2020gaussian} & CO, $NO_2$, $O_3$, $SO_2$, $PM_{10}$, and $PM_{2.5}$ data from six monitoring stations between March and April & Reduced-spaced Gaussian Process Regression and ANN \\ \hline
\cite   {sethi2020monitoring} & $CO$, $NO_2$, $O_3$, $SO_2$, $NH_3$, $PM_10$, $PM_{2.5}$, Toluene, benzene, and NH3 data from the Central Pollution Control Board and Ministry of Health and Family Welfare (MoHFW) & Decision tree (DT),   RF \\ \hline
\cite   {rahman2020air} & $PM_{2.5}$ data from monitoring station of the US Embassy, Dhaka, $SO_2$, $NO_2$, CO, and $O_3$ from AirNow, $NO_2$ measured by the Copernicus Sentinel-5 Precursor Tropospheric Monitoring Instrument & Generalized additive models (GAMs), wavelet coherence, RF \\ \hline
\cite   {tadano2020dynamic} & Daily COVID-19 cases and lockdown level from Statistical Portal of São Paulo State, and meteorological variables (e.g., relative humidity, maximum temperature, atmospheric pressure, wind speed, and global solar radiation), CO, $O_3$, $NO_2$, NO, $PM_{2.5}$, and $PM_{10}$ from Environmental Company of São Paulo State database (CETESB) & ANN models (Multilayer Perceptron overview, Radial basis function, Extreme Learning Machines, Echo State Networks) \\ \hline
\cite   {magazzino2020relationship} & The daily average of $PM_{2.5}$ and $PM_{10}$ from environmental monitoring stations located in the cities, COVID death, resuscitations, and hospitalization from the French National Public Health Agency & ANN \\ \hline
\cite   {mirri2020covid} & Italian Civil Protection, Regional Environmental Protection Agencies (ARPA) & SVM,  K-Nearest Neighbor (KNN), RF, GBDT, Classification and regression tree (CART), Multilayer perceptron (MLP), Ada boosting with decision tree (AdaBoost), Extra tree (ET) \\ \hline
\cite   {xu2020variation} & Nine observation sites in Hangzhou, China & RF \\ \hline
\cite   {cole2020impact} & City-level hourly data of 4 pollutants from Qingyue Open Environmental Data Center, meteorological data (e.g., temperature, relative humidity, wind direction, wind speed, and air pressure) from “worldmet” R package & RF, New augmented synthetic control method \\ \hline
\cite   {gatti2020machine} & COVID-19 positivity, mortality, and total case count from Italian Civil Protection, air pollutants (i.e., $PM_{2.5}$, $PM_{10}$, $NO_2$, $SO_2$, CO, Benzene, and $O_3$) from Italian Ministry of Agriculture, Food and Forestry and Regional Environmental Protection Agency (ARPA), air pollution from Italian National Institute of Statistics & RF \\ \hline
\cite   {guevara2020time} & Emission factors from different sources (open access and near-real-time measured activity data, proxy indicators   and other available reports), stringency index from Oxford COVID-19 Government Response Tracker (OxCGRT) & GBDT \\ \hline
\cite   {mele2020pollution} & Pollutant ($CO_2$, $PM_{2.5}$, and $NO_2$) emission from Open Government Data (OGD) and World Bank, per capita GDP from Federal Reserve Bank of St. Louis (FRED) & ML-based complex causality algorithm (D2C) \\ \hline
\cite   {vito2020high} & Pollutants (e.g., $NO_2$, CO, and $O_3$) data were captured by MONICA (a cooperative air quality monitoring station) and transmitted via a Bluetooth serial interface to a Raspberry Pi Mod. 3 + based datasink with Raspbian OS & Shallow   neural network (SNN) \\ \hline
\cite{wang2020four} & Pollutants (e.g., $NO_2$, $O_3$, $PM_{2.5}$, and CO) data from China National Environmental Monitoring Center, meteorological data (e.g., wind direction and speed,   temperature, relative humidity, and pressure) from NOAA Integrated Surface Database & RF \\ \hline
\cite   {barre2020estimating} & $NO_2$ data from operational Copernicus Sentinel 5 Precursor (S5P) TROPOMI, CAMS regional air quality models, European Centre for Medium-range Weather Forecasts (ECMWF) & GBDT \\ \hline
\end{tabular}
\end{table*}

\section{Conclusions, policy implications, and directions for future research}
\subsection{Conclusions}
The COVID-19 pandemic is one of the most devastating tragedies that humanity as a whole has experienced in the past centuries. It has adversely affected public health, social cohesion, health infrastructure, economic progress, transportation systems, and environmental quality. A range of preventive measures (e.g., lockdown and confinement measures, testing, vaccination), personal protection actions (e.g., face covering, hand washing), and financial assistance schemes (e.g., income support, debt relief) have been implemented to control the pandemic and reduce associated threats. As another dimension, the social distancing measures implemented nation-wide as a defensive response affect travel patterns of people and thereby influence air quality.

A considerable volume of studies have been conducted to predict COVID-19 transmission rates, evaluate the impacts of the pandemic on mobility and air quality, and assess the effectiveness of lockdown measures and the role of air quality on COVID-19 diffusion (Tables \ref{tab:geoCOntext} and  \ref{tab:airQGeoContext}). As a departure from pandemics that swept the world in previous decades, many of these studies have used ML techniques to understand the complex relationships between them (Tables \ref{tab:mobilityDataML} and \ref{tab:airQDataML}), leveraging the convergence of the widespread availability of huge volumes of structured and unstructured data enabled by wireless and mobile information technologies, and of ML techniques to analyze them with little to no restriction on distributional properties of data. Considering the distinctive ability of ML techniques to deal with multifaceted and wicked problems, this study aimed at reviewing the burgeoning body of past research that applied different ML tools to understand the intersecting relationships between the COVID-19 pandemic, lockdown measures, human mobility, and urban air quality.

Similar to other strands of studies (e.g., transportation, environment, and public sentiment), a large body of literature is using ML techniques in epidemiology research, which is a major breakthrough in medical history and has made significant contributions in transforming public healthcare systems \cite {deo2015machine, peng2020multiscale}. Many ML researchers have seized the opportunity to make their analytics relevant to the pressing needs of this time due to the convergence of data through cloud storage and data sharing via the interactive platform, and powerful analytic tools \cite {hicks2019best}. ML is very effective and well suited to handling notoriously wicked problems of epidemiology (i.e., multiscalarity, non-linearity, feedback effects, and so on), and it has proved itself worthwhile in previous research \cite {scavuzzo2018modeling, tai2019machine, alber2019integrating}. ML techniques predict outcomes with a higher rate of accuracy and reveal the hidden patterns in the data compared to traditional data processing systems. ML allows to integrate multiple modeling approaches to handle complex and large data sets contrary to conventional methods, which are more in line with the "one size fits all" principle. So, ML explicitly captures the context, whether spatial, social or environmental, as well as contextual changes; thus, predictions are adjusted accordingly, which is challenging to handle using traditional statistical and econometric approaches. Considering the enormous demand and potentials, the U.S. National Science Foundation (NSF) has initiated a series of transdisciplinary conversations in the form of webinars at the intersection of public health, behavioral and social sciences, smart technologies and data analytics to catalyze ideas that can further the research and development at the frontier of predictive intelligence for pandemic prevention \cite {nsf2021webinars}.

After critically analyzing the past studies on the emergence and evolution of the COVID-19 pandemic, this study observed the circumstances that amplify or mitigate the pandemic situation. In particular, factors of urban form, socioeconomic and physical conditions of the people, social cohesion, and social distancing measures significantly influence human mobility and stimulate the diffusion of the virus. Thus, social, physical, and institutional structures and the tendency of the people to follow social distancing measures sufficiently change travel patterns and determine the severity of the pandemic in a city or larger national territory. Human mobility and COVID-19 pandemic exhibit bidirectional relationships. During the COVID-19 period, people tend to travel less and are more likely to use private transportation for necessary travel purposes to mitigate coronavirus-related health problems. This review study also found that COVID-19 related lockdown measures significantly reduce the concentration of air pollutants and improve air quality due to reduced energy consumption in transportation and industries. This improved air quality also ameliorates the COVID-19 situation by reducing respiratory disease in the population. As a core approach to predictive analytics, ML is well positioned to bring newly learned inductive knowledge to the practice of pandemic prevention and management, but also to transform theories of pandemics through abductive inference.

\subsection{Policy implications}
Several policy implications can be drawn from the above discussion, as stated below. 
\begin{itemize}

    \item AI and ML techniques can be used in future multiscalar and non-linear disease modeling which can yield more accurate prediction results and formulate relevant premises for diseases diagnostic and prescribing treatment measures accurately \cite {tai2019machine}.
    
    \item Studies have found that COVID-19 related lockdown measures improved air quality by restricting human activities and altering the travel patterns. Thus, transportation policies (e.g., driving restriction on certain days or portion of the day, congestion pricing, emission standards for vehicles) to restrict auto driving, encourage public and active transportation, improve the fuel efficiency of the vehicles, and alternate working provisions, could be effective policy options to reduce air pollutants \cite {wang2020air, edf2020covid19, mori2020implications, cox2020environmental}. 
    
    \item Due to people's concerns on travel safety and hygiene, recent initiatives (e.g., dedicated lanes for bicycles and scooters), and changes in behaviors (e.g., teleworking, online shopping, and deliveries), mobility providers are re-evaluating the future of tomorrow’s mobility \cite {McKinsey2020mobility}. Policy makers should take initiatives to promote micro-mobility (e.g., cycling, e-scooters) and future transportation options (e.g., connected and autonomous vehicles, electric vehicles, shared mobility) should consider safety and safety issues to address the changes in consumer preferences. 
 
    \item Environmental policies (e.g., polluters pay, energy-efficient production, promotion of renewable energy) to restrict anthropogenic activities could reduce environmental pollution and respiratory illness in post-pandemic situations \cite {singh2020impact, feng2016legislation, cox2020environmental, rahman2018sustainability}. A comprehensive plan comprising of land use distribution, transportation, and environmental policies should be developed and continued to execute during and after the crisis, which are essential to achieve desired air quality standard \cite {cox2020environmental}. 

    \item Reducing people’s exposure to air pollutants through proper urban planning could be an effective policy option. For example, developing residential areas, healthcare facilities, educational institutions away from major sources of pollution (e.g., highway, industrial areas) are more likely to reduce exposure of people to the air pollutants \cite {edf2020covid19}. Moreover, city authorities could establish advanced pollutants monitoring stations and air filters particularly in the highly polluted areas to limit the emissions at the sources, decrease exposure to toxic air pollutants and protect public health \cite {edf2020covid19, apha2017airquality, wang2020changes}. 

    \item Outbreak and transmission of pathogenic diseases such as COVID-19 could be attributed to man-made and climate change-induced habitat alterations and interactions \cite {mori2020implications, cox2020environmental}. Thus, governments and policy makers across the globe should incorporate pandemic risk management strategies in climate action plans and adaptation schemes. 
    
    \item Considering the massive disruption of supply chains and the downfall of tourism sectors, policy makers should take appropriate initiatives locally and internationally to achieve a sustainable and resilient logistic management system and tourism industry \cite {mori2020implications}. 
    
\end{itemize}

\subsection{Limitations of past studies and directions for future research}
Despite the well-developed models we have on hand, convincing findings, and significant contributions to the literature in COVID-19 related mobility and air quality studies, the overall achievements of past studies are somewhat subdued by some recognizable reasons and hence there is scope for future research. We state some areas below. 

First, some studies used a limited number of variables to predict coronavirus cases and deaths, which may overlook other important factors that could influence disease transmission \cite {ayyoubzadeh2020predicting, iwendi2020covid, Lu2020CovidForcast}. Thus, future research should combine data from other sources such as social media, mass media, people’s contacts with a special call center for COVID-19, mobile contract tracing, environmental and climate factors, and screening registries \cite {ayyoubzadeh2020predicting}. Moreover, some researchers predicted COVID-19 cases for a limited period (i.e., only for the next 12 days), which seems insufficient to assess the performance of a model as a predictive tool \cite {pickering2020identifying}.

Second, some studies only investigated the impact of mobility patterns on the infection rate \cite {al2020measurement, Lu2020CovidForcast, shirvani2020correlation}. However, other factors (e.g., late government action, international travel, social structure, people's socioeconomic factors) may influence coronavirus cases and deaths. Thus, for accurate and meaningful prediction, these factors should be included in the models. Moreover, these studies mostly used mobility data (i.e., mobile device location based on GPS) from Google, Apple, and SafeGraph \cite {delen2020no, InteractiveCOVID19, kuo2020evaluating, hou2020intra, chiang2020hawkes, bao2020covid, roy2020characterizing}. These mobility reports provide inadequate information about people’s mobility since many of them may disable location settings in their mobile devices or may not have Android/iOS supported mobile. Moreover, some countries may not have reliable internet access and do not support Google (e.g., Iran, China), which may affect the quality and completeness of the data infrastructure. Researchers in \cite {chakraborty2020linear} suggested exploring additional information (e.g., vehicle miles traveled, number of miles traveled per person, trips per person, travel mode share) may explicitly consider those uncertainties to develop a more trusted model. 

Third, researchers are unable to determine an association between COVID-19 transmission and population characteristics in some high-incidence areas due to the difficulty to identify whether the cases have a local or international origin \cite {scarpone2020multimethod}. Moreover, the true number of incidence is unknown due to asymptomatic individuals, differences in testing and reporting, and misdiagnosis, which make the modeling very challenging. Some studies did not report the accuracy level, which call into question the validity of the model \cite {soures2020sirnet}. Thus, studies should incorporate some baselines (e.g., traditional fitting models) and other metrics for model evaluation (e.g., RMSE, summary statistics across bins, confusion matrix, cross-validation, etc.) \cite {bao2020covid}.

Fourth, some studies did not consider the impacts of seasonal variations on air quality \cite {tadano2020dynamic, velasquez2020gaussian, zambrano2020has, sethi2020monitoring, rahman2020air, magazzino2020relationship}. It is evident that air quality improves gradually from winter to summer. Therefore, there is conventional seasonal air quality. It is difficult to establish the effect of lockdown measures without considering the seasonal change. Thus, it is recommended to include seasonal variables in models to estimate the effects of lockdown measures on air quality to avoid any spurious and misleading relationships. 

Fifth, the impact of time traveling in shared mobility for coronavirus transmission is rarely studied in the literature. However, long duration travels in shared and crowded modes (e.g., air, bus, train, etc.) may transmit the contagious disease easily. Hence, this is a potential direction for future research. 

Sixth, despite huge data sharing during the COVID-19 pandemic, the application of AI and ML tools in epidemiology and other areas is still in its infancy due to limited access to reliable data infrastructure (i.e., platform, storage, network, collaboration). However, an efficient data infrastructure increases data acquisition, business collaborations, and operations by integrating data at various granularities and from diverse sources including unstructured data like social media data. Thus, a priority should be given to developing trustworthy data infrastructure to make data available to researchers for data-driven decision making.  

Seventh, a hybrid ML model (i.e., integration of multiple ML techniques) has enormous analytical ability to handle complex data and solve real-world problems such as COVID-19 compared to single ML techniques and simulation models. Thus, future studies should be directed to implement hybrid ML techniques for efficient and better performing models for real-world events \cite {tsai2013comparative, shao2014hybrid}.

\bibliographystyle{IEEEtran}

\bibliography{main.bbl}

\begin{IEEEbiography}[{\includegraphics[width=1in,height=1.25in,clip,keepaspectratio]{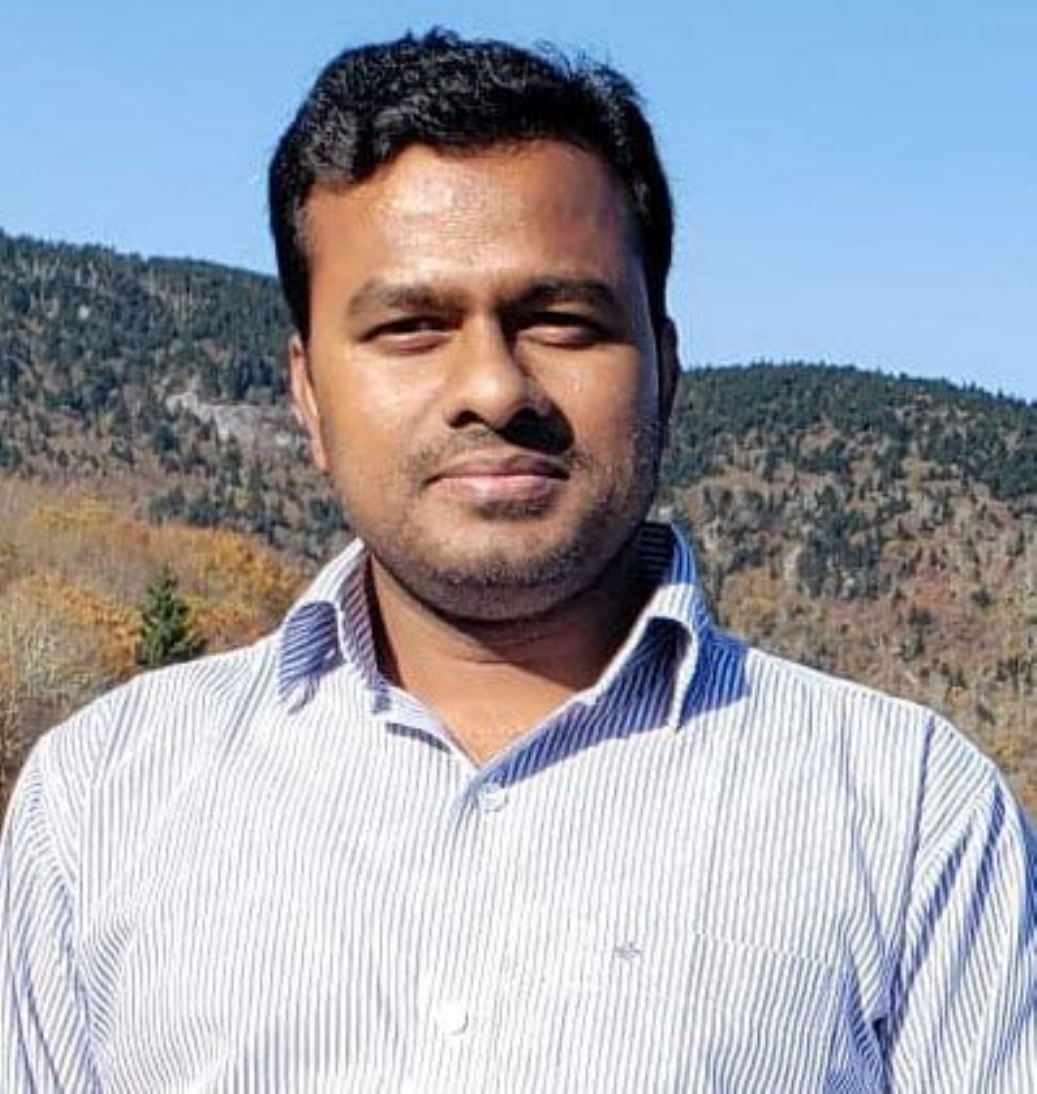}}]{Md. Mokhlesur Rahman} has completed his Bachelor of Urban and Regional Planning (B.URP) from the Department of Urban and Regional Planning (URP), Bangladesh University of Engineering \& Technology (BUET) in 2009. To compete in the professional arena and acquire advanced knowledge, he has completed an M.Sc in Urban Planning from the Department of Urban Planning and Design (DUPAD), University of Hong Kong (HKU), Hong Kong. After graduation from HKU, he started working at Khulna University of Engineering \& Technology (KUET) as a Lecturer in December 2014 and promoted to Assistant Professor in February 2017. Currently, he is a graduate student and research assistant at the University of North Carolina at Charlotte, North Carolina, USA. His research interests include Transportation Planning and Engineering, Travel Behaviors and Demand Modeling, Land Use Modeling, Climate Change and Disaster Management, Application of GIS in Transportation and Environment, Econometric Modeling, and Data Analytics.
\end{IEEEbiography}

\begin{IEEEbiography}[{\includegraphics[width=1in,height=1.25in,clip,keepaspectratio]{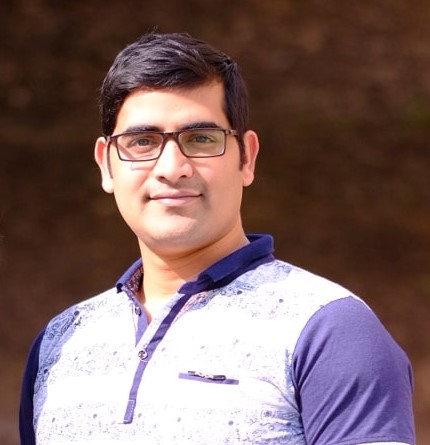}}]{Kamal Chandra Paul} is a PhD student at the department of Electrical and Computer Engineering, the University of North Carolina at Charlotte (UNCC). He received his bachelor degree in Electrical and Electronic Engineering (EEE) from Khulna University of Engineering \& Technology (KUET) in 2009. He obtained his MS degree in Electrical Engineering from the Ingram School of Engineering, Texas State University, USA in 2018. He served as a faculty at the department of EEE, International University of Business Agriculture and Technology (IUBAT). He also worked as a lecturer in the department of EEE at World University of Bangladesh. His research interest includes AC and DC arc fault detection, data science, application of ML in power electronics, and renewable energy.
\end{IEEEbiography}
\vspace{-1 cm}  

\begin{IEEEbiography}[{\includegraphics[width=1in,height=1.35in,clip,keepaspectratio]{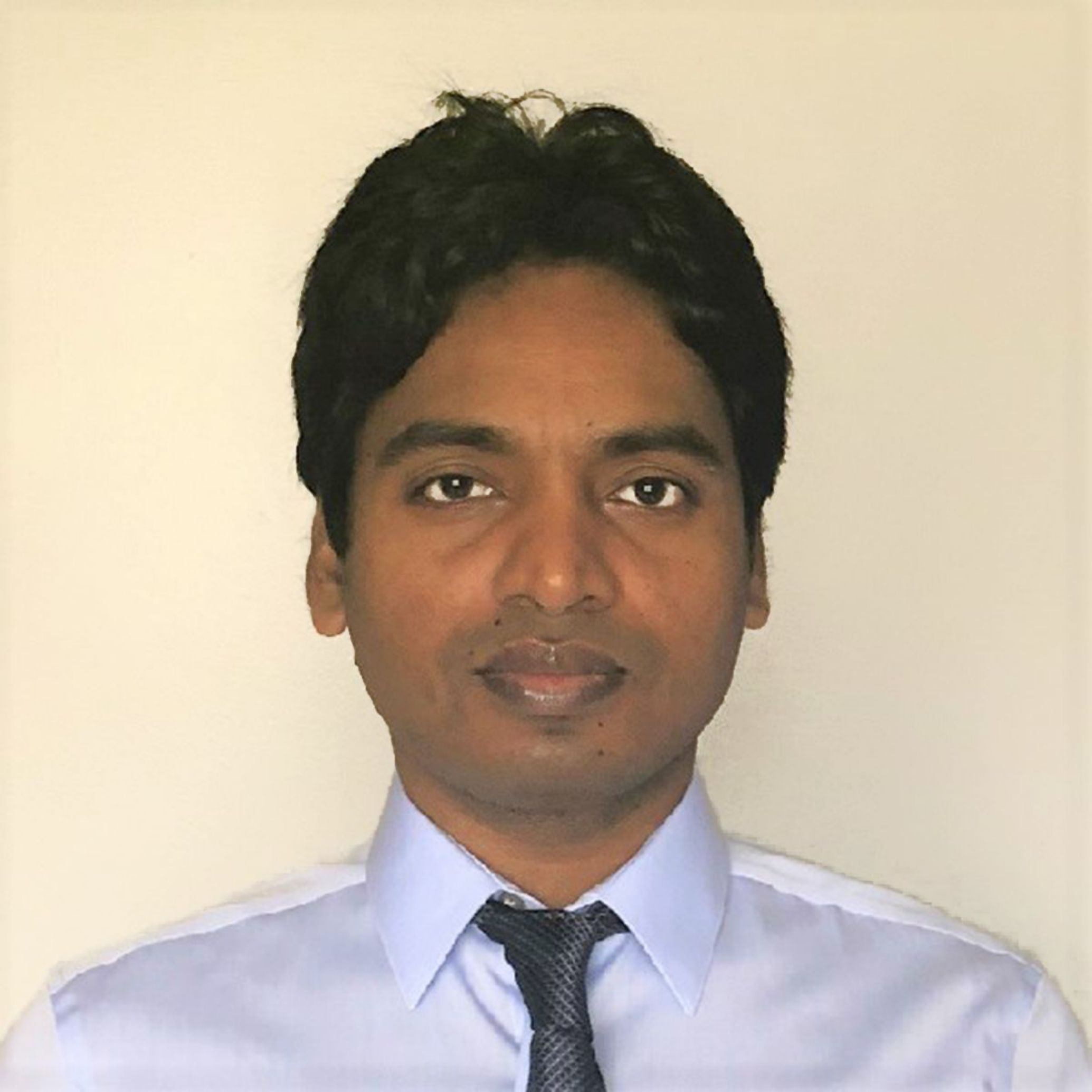}}] {Md. Amjad Hossain} received B.Sc. degree in computer science and Engineering from Khulna University of Engineering and Technology(KUET), Bangladesh, in 2008. He completed his Ph.D. in Computer Science from Kent State University, USA, in 2020. Hossain started working as a Lecturer in KUET immediately after his B.Sc. Later, he received a promotion to Assistant Professor and continued working in KUET until 2012. Currently, he is working as an Assistant Professor of computer and information sciences at Shepherd University, WV, USA. His current research interest includes distributed systems, multimedia computing and networking, ML, and Image Processing.  
\end{IEEEbiography}

\begin{IEEEbiography}[{\includegraphics[width=1in,height=1.25in,clip,keepaspectratio]{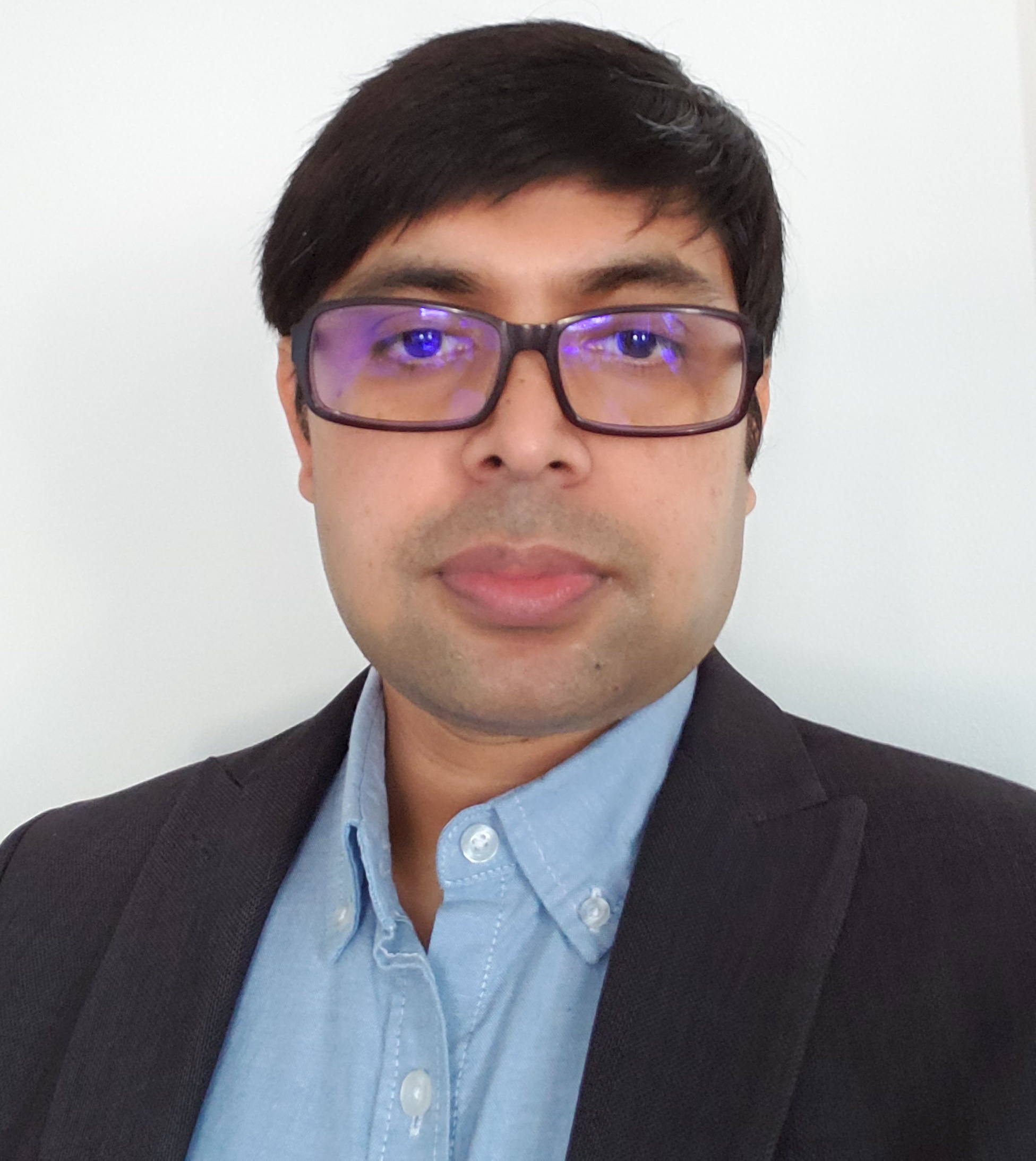}}]{G. G. Md. Nawaz Ali} (M'15) received his B.Sc. degree in Computer Science and Engineering from the Khulna University of Engineering \& Technology, Bangladesh in 2006, and the Ph.D. degree in Computer Science from the City University of Hong Kong, Hong Kong in 2013 with the Outstanding Academic Performance Award. He is currently working as an Assistant Professor with the Department of Applied Computer Science of the University of Charleston, WV, USA. Prior to joining UCWV, he was a Post-doctoral Fellow with the Department of Automotive Engineering, The Clemson University International Center for Automotive Research (CU-ICAR), Greenville, SC, USA from March 2018 to July 2019. From October 2015 to March 2018, he was a postdoctoral research fellow with the School of Electrical and Electronic Engineering of Nanyang Technological University (NTU), Singapore. He is a reviewer of a number of international journals including the IEEE TRANSACTIONS ON INTELLIGENT TRANSPORTATION SYSTEMS AND MAGAZINE, IEEE TRANSACTIONS ON VEHICULAR TECHNOLOGY, IEEE INTERNET OF THINGS (IOT) JOURNAL, IEEE ACCESS, AND Wireless Networks etc. His current research interests include Vehicular Cyber Physical System (VCPS), wireless broadcasting, mobile computing, and network coding.
\end{IEEEbiography}
\vspace{-1 cm}  

\begin{IEEEbiography}[{\includegraphics[width=1in,height=1.25in,clip,keepaspectratio]{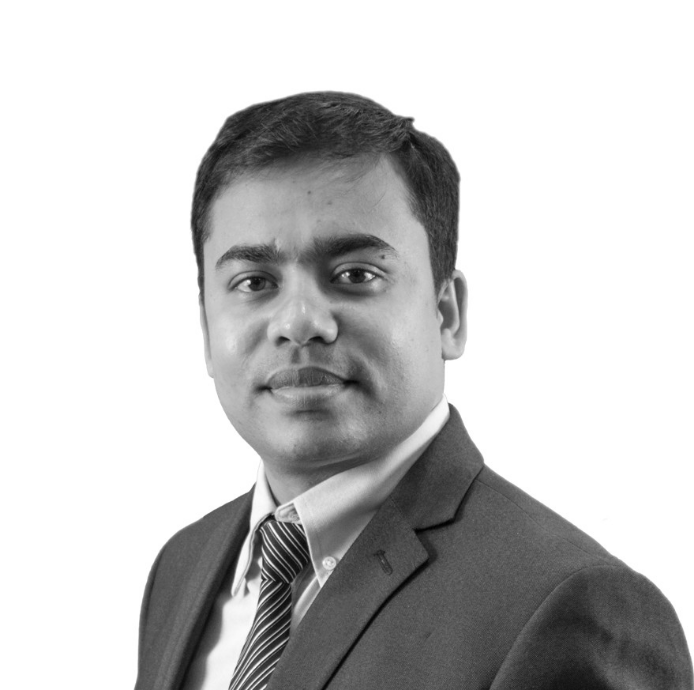}}]{Md. Shahinoor Rahman} is an assistant professor at the Department of Earth and Environmental Sciences of New Jersey City University (NJCU). He completed his Ph.D. in Earth Systems and Geoinformation Sciences from George Mason University (GMU) in December 2019. He is the recipient of the Outstanding Ph.D. Student Award 2020 from the Department of Geography and Geoinformation Science at Mason for his excellence in the studies, prolific track record of scholarly publications. A former Assistant Professor of the apex engineering university in Bangladesh, Bangladesh University of Engineering and Technology (BUET), he completed his undergraduate in Urban and Regional Planning from BUET and masters in Regional Development Planning and Management jointly offered by TU-Dortmund, Germany and Universidad Austral de Chile. He primarily specializes in the application of remote sensing (RS) and geographic information science (GIS) to examine the cause of–and innovative solutions to–some of our world’s most pressing environmental challenges including climate change impact on soil and water, agricultural damage, coastal vulnerability, natural hazard risks, environmental degradation by urban expansion, and urban heat island effect. He utilizes data mining, ML, and spatial analytics approaches in his research. His broad goal is to utilize advanced geospatial technologies for the betterment of our society and environment. 
\end{IEEEbiography}
\begin{IEEEbiography}[{\includegraphics[width=1in,height=1.25in,clip,keepaspectratio]{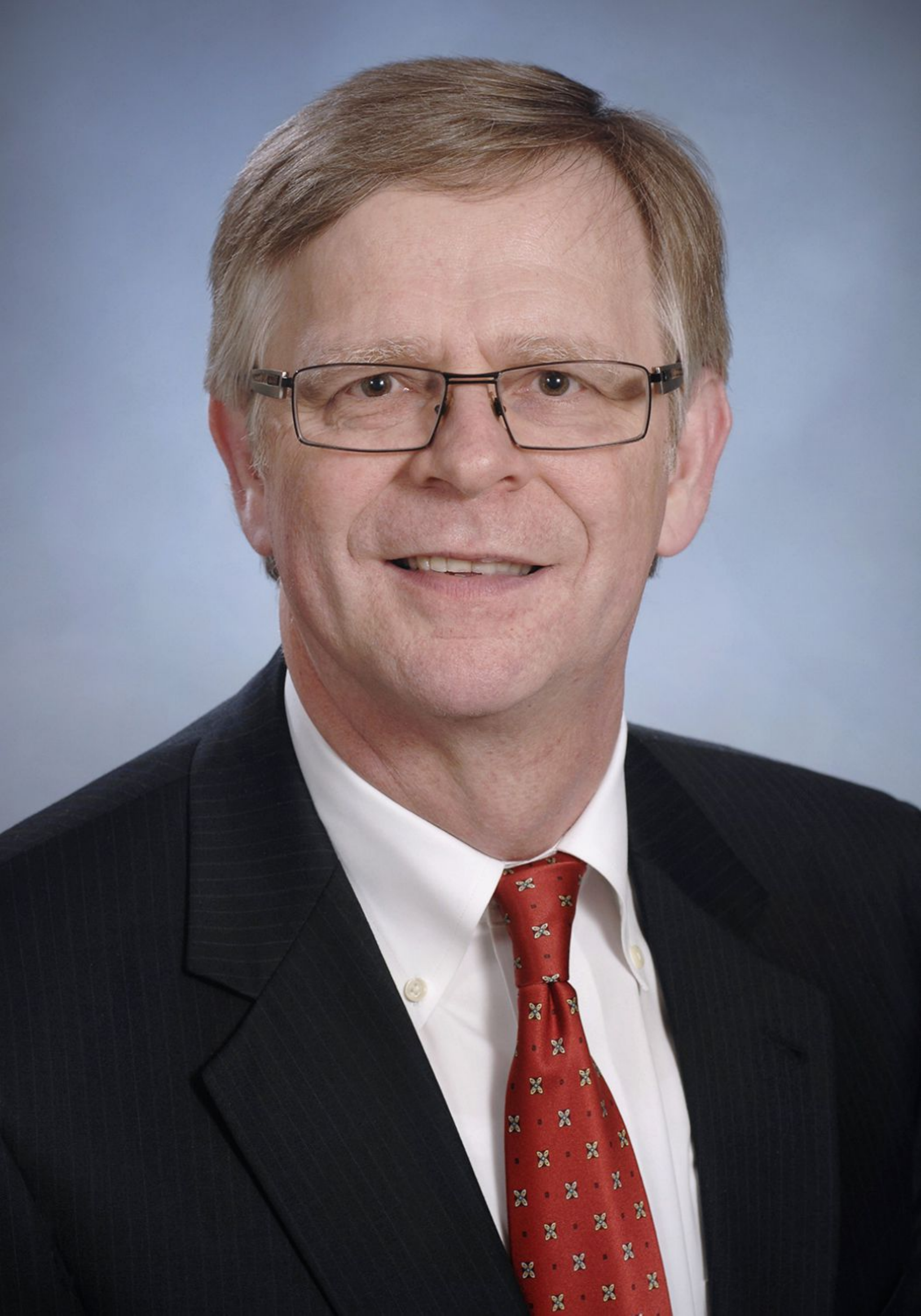}}]{Jean-Claude Thill} is a Knight Foundation Distinguished Professor of Public Policy in the Department of Geography and Earth Sciences at the University of North Carolina at Charlotte. He is an Affiliate Faculty of the School of Data Science and holds graduate degrees from the Catholic University of Louvain, Belgium. Dr. Thill is an urban economic geographer whose research focuses on urban and regional transportation and mobility issues, processes of urbanization, regional science, and geospatial data science. His recent research focuses on modeling past urbanization trajectories of the Charlotte metropolitan area (land use transformation, local and regional drifts in neighborhood quality of life), livability, innovation, and mobility in Charlotte, other US cities, and Chinese cities, and accessibility to public services. He has more than 30 years of teaching and research experience in transportation planning and spatial modeling. His current and recent grants have been awarded from the National Science Foundation, Department of Defense, Department of Energy, North Carolina Department of Transportation, World Bank, and more. He is the 2012 recipient of the Edward L. Ullman Award for Significant Contributions to Transportation Geography, awarded by the American Association of Geographers. He has also received the Hirotada Kohno Award for Outstanding Service of the Regional Science Association International in 2012. Since 2013, he is a Fellow of the Regional Science Association International. He is honored to have been appointed High-End Foreign Expert in the School of Applied Economics, Renmin University of China (Ministry of Science and Technology), in Beijing, P.R. China.
\end{IEEEbiography}
\end{document}